\documentclass[conference]{IEEEtran}

\makeatletter
\renewcommand*{\@fnsymbol}[1]{\ensuremath{*}}
\makeatother
\newcommand*\samethanks[1][\value{footnote}]{\footnotemark[#1]}
\newcommand{\E}{\mathop{\mathbb E\/}}

\usepackage[T1]{fontenc}
\usepackage[utf8]{inputenc}
\usepackage{pslatex}
\usepackage{multicol}
\usepackage{xcolor}
\usepackage{textcomp} 

\usepackage[kerning,spacing]{microtype} 
\usepackage{flushend} 

\usepackage{cite}               
\usepackage{url}                
\usepackage{url}

\usepackage{breakurl}
\usepackage[breaklinks]{hyperref}

\usepackage{xcolor}             
\usepackage[]{hyperref}         
\hypersetup{                    
  colorlinks,
  linkcolor={green!80!black},
  citecolor={red!70!black},
  urlcolor={blue!70!black}
}

\usepackage{breakurl}           

\usepackage{amsmath,amsthm}
\usepackage{cleveref}
\usepackage{booktabs,multirow,array,caption}
\usepackage{stackengine}
\usepackage{comment}

\usepackage{tikz}

\usepackage{filecontents}

\usepackage[utf8]{inputenc}

\usepackage{enumitem}
\usepackage{bbm}
\usepackage{amsmath,amssymb,amsthm}
\usepackage{xspace,graphicx}
\usepackage{diagbox}

\newtheorem{thm}{Theorem}[section]
\newtheorem{prop}[thm]{Proposition}

\newcommand{\Cite}[1]{\protect{\cite{{#1}}}}

\newcommand{\sprof}{S_{prof}}

\newcommand{\name}{SquirRL}
\newcommand{\honestagent}{\ensuremath{\mathcal{H}}\xspace}
\newcommand{\stratagent}{\ensuremath{\mathcal{S}}\xspace}

\newcommand{\sys}{{\sf Sys}}
\newcommand{\mechanism}{\ensuremath{\mathcal{M}}\xspace}

\title{\name: Automating Attack Analysis on Blockchain Incentive Mechanisms with Deep Reinforcement Learning}
\date{}

\author{
{\rm Charlie Hou\thanks{Equal contribution}}\\
Carnegie Mellon University, IC3
\and
{\rm Mingxun Zhou\samethanks}\\
Peking University
\and
{\rm Yan Ji}\\
Cornell Tech, IC3
\and
{\rm Phil Daian}\\
Cornell Tech, IC3
\and
{\rm Florian Tram\`er}\\
Stanford University
\and
{\rm Giulia Fanti}\\
Carnegie Mellon University, IC3
\and
{\rm Ari Juels}\\
Cornell Tech, IC3
}

\makeatletter
\newcommand{\linebreakand}{%
  \end{@IEEEauthorhalign}
  \hfill\mbox{}\par
  \mbox{}\hfill\begin{@IEEEauthorhalign}
}
\makeatother

\author{
\IEEEauthorblockN{Charlie Hou\IEEEauthorrefmark{1}} 
\IEEEauthorblockA{\textit{Carnegie Mellon University, IC3} \\
charlieh@andrew.cmu.edu}
\and
\IEEEauthorblockN{Mingxun Zhou\IEEEauthorrefmark{1}} 
\IEEEauthorblockA{\textit{Peking University} \\
zhoumingxun@pku.edu.cn}
\and
\IEEEauthorblockN{Yan Ji}
\IEEEauthorblockA{\textit{Cornell Tech, IC3} \\
yj348@cornell.edu}
\and
\IEEEauthorblockN{Phil Daian}
\IEEEauthorblockA{\textit{Cornell  Tech, IC3}\\
pad242@cornell.edu}
\linebreakand
\IEEEauthorblockN{Florian Tram\`er}
\IEEEauthorblockA{\textit{Stanford University} \\
tramer@cs.stanford.edu}
\and
\IEEEauthorblockN{Giulia Fanti}
\IEEEauthorblockA{\textit{Carnegie Mellon University, IC3} \\
gfanti@andrew.cmu.edu}
\and
\IEEEauthorblockN{Ari Juels}
\IEEEauthorblockA{\textit{Cornell Tech, IC3} \\
juels@cornell.edu}
}

\hypersetup{draft}
\begin{document}

\maketitle



\begin{abstract}
 Incentive mechanisms are central to the functionality of permissionless blockchains: 
 they incentivize participants to run and secure the underlying consensus protocol. 
Designing incentive-compatible incentive mechanisms is notoriously challenging, however. 
As a result, most public blockchains today use 
incentive mechanisms whose security properties are poorly understood and largely untested. 
In this work, we propose \name, a framework for using deep reinforcement learning to analyze
attacks on blockchain incentive mechanisms.  
We demonstrate \name's power by first recovering known attacks: (1) the optimal selfish mining attack in Bitcoin \cite{sapirshtein2016optimal}, and (2) the Nash equilibrium
in block withholding attacks \cite{eyal2015miner}. 
We also use \name~to 
obtain several novel empirical results.  
First, we discover a counterintuitive flaw in the widely used \emph{rushing adversary} model when applied to multi-agent Markov games with incomplete information.  Second, we demonstrate that the optimal selfish mining strategy identified in \cite{sapirshtein2016optimal} is actually not a Nash equilibrium in the multi-agent selfish mining setting.  In fact, our results suggest (but do not prove) that when more than two competing agents engage in selfish mining, there \emph{is no profitable Nash equilibrium}. This is consistent with the lack of observed selfish mining in the wild. Third, we find a novel attack on a simplified version of Ethereum's finalization mechanism, Casper the Friendly Finality Gadget (FFG) that allows a strategic agent to amplify her rewards by up to $30\%$.  
Notably, \cite{buterin2019incentives} shows that honest voting is a Nash equilibrium in Casper FFG; our attack shows that when Casper FFG is composed with selfish mining, this is no longer the case. 
Altogether,
our experiments demonstrate \name's flexibility and promise as a framework for studying attack settings that 
have thus far eluded theoretical and empirical understanding.

\end{abstract}

\section{Introduction}
\label{sec:intro}

Blockchains today require participants to expend substantial resources  (storage, computation, electricity) to ensure the correctness and liveness of other users' transactions. 
Most public blockchains therefore rely critically on \emph{incentive  mechanisms} to motivate users to participate in blockchain consensus protocols.
That is, users are typically paid (in native cryptocurrency) to sustain the system. 
Incentive mechanisms are therefore central to the functionality of most permissionless blockchains. 

For example, Bitcoin's consensus protocol requires  participants  (also  known  as   \emph{miners}) to build a sequential data structure of \emph{blocks}, where each block is generated in a computationally intensive process (called \emph{mining}).  
To incentivize participation, Bitcoin miners receive a \emph{block reward} (in Bitcoins) for every block they mine that is accepted by the rest of the network. 
Miners also receive smaller \emph{transaction fees} for  transactions they include in a block; this is done to prevent miners from simply mining empty blocks without doing useful work.
The incentive mechanisms of block rewards and fees have driven the remarkable growth of the Bitcoin ecosystem.


In practice, poorly-designed incentive mechanisms can be exploited by rational users. By deviating from the protocol-specified behavior, users may be able to accumulate more rewards than what they are nominally entitled to.
For example, \emph{selfish mining} is a well-known attack on Bitcoin's incentive mechanism that allows a strategic miner to reap more than her fair share of block rewards by waiting to publish her blocks until she would cause the most damage to the honest majority  \protect{\cite{eyal2018majority}}. 
Many subsequent papers have explored both attacks on Bitcoin's incentive mechanism \protect{\cite{eyal2015miner,sapirshtein2016optimal,nayak2016stubborn,carlsten2016instability,judmayer2019pay,liao2017incentivizing}} as well as other cryptocurrencies \protect{\cite{ritz2018impact,grunspan2019selfish,niu2019selfish,neuder2019selfish}}.
%

Today, attacks on blockchain incentive mechanisms are typically studied through a lengthy process of modeling and theoretical analysis \protect{\cite{eyal2015miner,sapirshtein2016optimal,nayak2016stubborn,carlsten2016instability,judmayer2019pay,liao2017incentivizing,ritz2018impact,grunspan2019selfish,niu2019selfish}}.  
As many cryptocurrencies lack the resources for theoretical analysis, the vast majority of blockchain incentive mechanisms have not been studied at all. Substantial amounts of cryptocurrency may thus be vulnerable to unknown attacks.


\subsection{Proposed framework: \name}
In this work, we propose \name,  a generalizable  framework for using deep reinforcement learning (DRL) to analyze blockchain incentive mechanisms.
\name~is intended as a general-purpose methodology for blockchain developers to test incentive mechanisms for vulnerabilities. 
It does not provide theoretical guarantees: just because it does not find any profitable attacks  does not mean that honest behavior is a dominant strategy. 
We find in  practice, however, that instantiations of \name~are effective at identifying adversarial strategies, which can be used to prove that an incentive mechanism  is \emph{insecure}.
Our primary contributions are threefold:
\begin{enumerate}[leftmargin=0cm,itemindent=.5cm,labelwidth=\itemindent,labelsep=0cm,align=left]
\item {\em Framework:} We present~\name~as a a general framework for exploring vulnerabilities in blockchain incentive mechanisms and recovering adversarial strategies. The framework broadly involves: (1) creation of a simulation environment, with  accompanying feature  
and action spaces reflecting the views and capabilities of participating agents; (2) selection of an adversarial model, including numbers and types of agents; and (3) selection of a suitable RL algorithm and associated reward function. 
As part of this framework, we develop a general state space representation for a broad class of blockchain incentive mechanisms, which allows us to trade off feature dimensionality with accuracy.  
We show how to use this framework flexibly to handle settings involving varying environments, numbers of agents, and rewards.

\item {\em Selfish-mining evaluation:} We apply \name~to various blockchain consensus/incentive protocols to analyze variants of selfish mining. 
Using \name, we are able to recover known theoretical results in the Bitcoin protocol, while also extending state-of-the-art results to domains that were previously intractable (e.g., the multi-agent setting, larger state spaces, other protocols).  
Our experiments suggest two new findings: 
\begin{itemize}[align=left]
	\item Theoretically analyzing repeated interactions between multiple strategic agents is difficult due to the large state and action space. Prior work in this setting has therefore simplified the strategy space \cite{semiselfish}, under which a  \emph{semi-selfish mining} strategy is shown to be a Nash equilibrium.
    However, our experiments suggest that under a more general strategy space, semi-selfish mining is not a Nash equilibrium.
	In fact, for the Bitcoin protocol, \emph{all variants of selfish mining} appear to be unprofitable in settings with at least three strategic agents. This is consistent with the lack of observed selfish mining in the wild, although it is unclear whether this observation or other externalities are to blame.
	\item We find that the classical notion of a \emph{rushing adversary}, which is widely used in the cryptographic literature to model a worst-case adversary \cite{rushingex,rushingex2}, can give counterintuitive and nonphysical results in multi-strategic-agent settings. This has implications beyond the blockchain domain regarding how security researchers should model multi-agent settings. We expect this observation may be particularly relevant as DRL gains adoption as a tool for learning to attack and/or defend complex systems that are not amenable to theoretical analysis \cite{nguyen2019deep,wang2019deep,sengupta2020multi}. 
\end{itemize}

%
\item {\em Demonstration of extensibility:} We show that \name~is generally applicable to attacks on other types of incentive mechanisms beyond selfish mining:
\begin{itemize}[align=left]
	\item We apply \name{} to the proposed Ethereum finalization mechanism, Casper the Friendly Finality Gadget (FFG) \cite{casper}. These results illustrate that a strategic miner can collude with a Casper FFG validator to amplify its rewards by up to 30\% more than a strategic miner or validator could do alone. Such strategic collusion can cause honest validators to leave, progressively corrupting the system.  This raises important questions about the \emph{composability} of incentive mechanisms. 
	\item We apply \name{} to block withholding attacks, where it  converges to two-player strategies that match the Nash equilibrium in \cite{eyal2015miner}. 
\end{itemize}
\end{enumerate}

\paragraph{Paper outline}
We motivate the need for~\name~in \S\ref{sec:motivation},  explaining why existing techniques fall short, and then
provide background on  deep reinforcement learning in \S\ref{sec:rl}. We present the design  of  \name~in \S\ref{sec:system}.
We evaluate \name~on a variety of settings: the single-strategic-agent selfish mining setting in \S\ref{sec:eval}, the multi-strategic-agent selfish mining setting in \S\ref{ssec:eval_multi}, and Casper FFG and the Miner's Dilemma in \S\ref{sec:case-studies}.
We discuss related work in~\S\ref{sec:related} and conclude with a brief summary in~\S\ref{sec:conclusion}.

\section{Motivation} 
\label{sec:motivation}
Today, the process for analyzing new  attacks on blockchain incentive mechanisms is manual and time-consuming. 
Typically, studying such attacks require some combination of theoretical analysis, simulation, and intuition \protect{\cite{sapirshtein2016optimal}}.
Each becomes more difficult to obtain as the complexity of the underlying protocol grows.
In  particular, game-theoretic analysis of these systems tends to be challenging  for three reasons: (1) the state space is large (or continuous), (2) the game is repeated, and (3) there can be many agents.
Indeed, much of the existing analysis  has focused on settings where only one or two agents are allowed to deviate from the honest mining strategy  \protect{\cite{sapirshtein2016optimal,semiselfish}}.

At the same time, new protocols are emerging frequently,  each  with its own incentive mechanism \cite{blackcoin,particl,polkadot}. 
Oftentimes, protocol designers rely on intuition to reason about the security of their incentive mechanisms, in part because we lack general-purpose tools for mechanism analysis. 
Even when protocol designers provide security proofs, they typically at most show that honest behavior is a Nash equilibrium if honest parties are a signficiant portion of the participants in the protocol \protect{\cite{casper, pass2017fruitchains}}.
This weak guarantee says nothing about other equilibria or the (perhaps more realistic) setting in which many competing parties behave rationally.
Our central premise is that a systematic and largely automated approach for testing incentive mechanisms would streamline this process, and could help catch incentive mechanism bugs \emph{before} deployment in the wild. 

\begin{center}
\begin{table*}
\centering
\resizebox{\textwidth}{!}{
 \begin{tabular}[width=\textwidth]{@{}c   c  c  l@{}} 
{\bf Number of Strategic Agents}
& {\bf Representative Setting} & 
{\bf Agent Types} &
\hspace{5mm}{\bf Explored questions} \\
 \toprule
 1 & Single strategic agent & 
 $\sys \rightarrow \stratagent$ & $\begin{array}{l}
 \textrm{\textbullet\hspace{2mm}What impact from worst-case attack?}\cr
 \textrm{\textbullet\hspace{2mm}What is optimal adversarial strategy?} 
\end{array}$
 \\
 \midrule
 2 & 
 $\begin{array}{c}\cr
 \textrm{Emergent strategic-agent}\cr
 \textrm{behavior}
\end{array}$
 & \stratagent vs. \sys &
 $\begin{array}{l}
 \textrm{\textbullet\hspace{2mm}Is $\stratagent$ dominant?}\cr
 \textrm{\textbullet\hspace{2mm}Is $\stratagent$ profitable for competing agent?} 
\end{array}$
 \\
\cmidrule{3-4}
 && \sys~vs. \sys & $\begin{array}{l}
 \textrm{\textbullet\hspace{2mm}Is two-agent game stable?} 
\end{array}$
\\
 \midrule
$k \geq 3$ & 
 $\begin{array}{c}
 \textrm{Community of competing}\cr
 \textrm{strategic agents} 
\end{array}$
&  $\underbrace{\sys \text{~vs.~} \ldots \text{~vs.~}\sys}_{k \text{ agents}}$ & $\begin{array}{l}
 \textrm{\textbullet\hspace{2mm}Is multi-agent strategic play profitable?} \cr
  \textrm{\textbullet\hspace{2mm}Is \honestagent dominant in multi-agent setting?}
\end{array}$ \\ [1ex] 
 \bottomrule
\end{tabular}}
\caption{Experimental progression in an automated incentive mechanism analysis system \sys. The sequence of experiments with increasing numbers of strategic agents sheds light on key security questions for mechanism $\mechanism$. Notation $\sys \rightarrow \stratagent$ is shorthand denoting $\stratagent$ as the output of the automated system.}
\label{table:experimental_progression}
\end{table*}
\end{center}


\vspace{-8mm}

\subsection{Use Case} 
We envision protocol designers using our framework to study a natural progression of adversarial models and experiments to help address key security and incentive-alignment questions for an incentive mechanism $\mechanism$. 
These are shown in Table~\ref{table:experimental_progression}.

For a given adversarial resource (e.g., mining power), a {\em single} strategic agent $\stratagent$ competing against a group of honest agents represents the most powerful possible adversary.
(Such an agent can simulate any set of strategies among multiple agents.) 
$\stratagent$ can steal rewards from honest agent $\honestagent$, whose strategy is fixed {\em a priori} and thus cannot develop a counter-strategy.  
Learning $\stratagent$ thus yields insights into the worst-case performance of \mechanism. 

Addition of a second strategic agent then addresses the question of whether $\stratagent$ itself is dominant or suboptimal in the presence of a competing agent. This setting captures the dynamics when a single strategic agent is first challenged by others. By training two competing agents in tandem, it is also possible to explore questions such as:
How stable are learned strategies over time?

In actual deployment of a mechanism \mechanism, of course a community of $k \geq 3$ competing strategic agents can arise, a case more plausible than sustained attack by a single strategic agent. 
The~\name~framework dictates analysis with various values of $k$ to explore the likely {\em practical} security of \mechanism. For example, a mechanism may have poor worst-case security yet have its participants converge to the strategy of $\honestagent$ for all players given competition among strategic agents. A key question is: How much reward, as a function of $k$, can  strategic agents collectively steal from \honestagent?
We emphasize that the experiments in Table~\ref{table:experimental_progression} are a starting point, not a full prescription.
However, they shed light on a number of central, incentive-related questions that are nontrivial to evaluate today.

\subsection{Straw-man solution}
A natural first step for analyzing consensus protocols is modeling them as {Markov Decision Processes} (MDPs), and using classical algorithms such as \textbf{policy iteration} or \textbf{value iteration} \cite{csababook} to solve them.
MDPs are commonly used to model problems where an agent wishes to maximize its reward in a known, random environment \protect\cite{white2001markov}. 
Value iteration and policy iteration have been used effectively to computationally learn  optimal adversarial strategies in the two-agent (one strategic, one honest) setting of the Bitcoin protocol \protect{\cite{sapirshtein2016optimal}}.

MDPs are defined as a tuple $(S,A, P, R)$, where $S$ denotes a set of states, $A$ denotes a set of actions the agent can take, $P$ denotes the probability transition matrix, where $P_a(s,s')=\mathbb P(s_{t+1} = s'|s_t=s, a_t=a)$ denotes the probability of the agent transitioning  to state $s'$ from state $s$ by taking action $a$. $R$ is the reward matrix,  where $R_a(s,s')$ denotes the expected reward associated with transitioning from  state $s$ to $s'$ by taking action $a$.
We highlight one aspect of this definition. 
It relies critically on a Markov assumption, which states that the probability distribution over states depends only on the previous state and the action taken  at each  time step.
Conditioned on these assumptions, the objective in an MDP is to recover a strategy $\pi$ that optimizes the expected discounted long-term reward $\mathbb E[\sum_{t=0}^\infty \eta^{\,\,t} R_{a_t}(s_t,s_{t+1})]$, where $\eta  \in (0,1)$ is a discount factor that accounts for how much the agent values short-term rewards over long-term ones, 
and the expectation is taken over the randomness in the system evolution and the potentially randomized strategy $\pi$.

With known and exactly specified $(S,A,P,R)$, value iteration or policy iteration can exactly solve (up to desired precision) for the optimal $\pi$.  
In \protect{\cite{sapirshtein2016optimal}}, policy iteration was used to find an optimal selfish mining strategy for a rational Bitcoin agent.
However, policy and value iteration exhibit two primary constraints that prevent them from being a useful general-purpose tool  for our problem.

\emph{1. They assume a stationary environment.} 
To formulate an MDP, there must exist a fixed probability transition  matrix $P$.
This is true in the single-strategic-agent setting where one agent is honest and thus behaves according to a known strategy. 
However, in practical settings, we may have multiple rational agents who are dynamically changing their strategies, leading to a non-stationary environment. 
This is no longer an MDP, but a Markov game, where policy iteration and value iteration do not apply.

\emph{2. They scale poorly with growing state spaces.} 
Policy and value iteration store the probability transition matrix $P$ and reward matrix $R$ explicitly, which requires storage $O(|S|^2 |A|)$ (a probability value must be stored for each transition $(s,a,s')$, where $s$ is the current state, $a$ is the action taken, and $s'$ is the next state).    
This can be prohibitive for protocols where the state cannot be represented by a compact feature, e.g., when the reward is not computed from a single chain in the ledger's directed acyclic graph (DAG)  \cite{wood2014ethereum,ghost,conflux}.

Even in Bitcoin, the state space can be intractably large.  For example, if there are two strategic (selfish mining) agents A and B, agent A cannot observe the hidden blocks of agent B.  Agent A needs to at least have an unbiased estimate of B's hidden blocks to write out an MDP.  The only way for agent A to estimate agent B's hidden blocks is to use the history of its observations as well as some notion of time.  Suppose that we wanted to consider a past history of $t_0$ observations, and the upper limit for our feature for time is $T$.  Then we need storage $O((|S|T)^{t_0 + 2} |A|)$ (where we let $S$ be the space of all Bitcoin blocktrees), which quickly grows infeasible.


\section{Deep Reinforcement Learning}
\label{sec:rl}
Reinforcement learning (RL) is a class of machine learning algorithms that learn strategies enabling an agent to maximize its cumulative rewards in an environment.  Much RL research is focused on solving MDPs when $P$ or $R$ are unknown or too large to be practical \cite{csababook}.  However, the field encompasses more general settings, such as Markov games.  
 \emph{Deep reinforcement learning (DRL)} is a class of RL that uses neural networks to learn policies, often without needing to explicitly specify the underlying system dynamics \cite{franccois2018introduction}. 
In this work, we explore the potential of deep reinforcement learning to automate the analysis of attacks on blockchain incentive mechanisms, particularly in settings where algorithms like policy iteration are impractical or impossible to use (e.g., large or infinite state spaces, Markov games, and MDPs with unspecified $P$ or $R$).  

DRL has been particularly successful in problems where the state space is intractably large.  Roughly, this is because DRL uses neural networks to replace tables; for instance, instead of storing a lookup table to decide what action to take at each $s \in S$, one can instead have a function (neural network) $f: S \to A$ whose size does not scale with $|S|$.     
Widely-publicized examples like chess and Go exhibit large state spaces \protect{\cite{alphago,chess}}, 
as do blockchain incentive mechanisms. 

Most blockchain incentive mechanisms have an additional RL-friendly feature in that rewards are processed continuously. 
That is, in chess or Go, rewards are calculated in an  all-or-nothing manner at the end of the game. 
In blockchains, players reap rewards incrementally, enabling reward estimation before the game is complete. This faster feedback makes it easier to train automated systems to learn effective strategies.
As such, DRL is a natural tool for this problem.


\subsection{Design considerations in deep RL}

\textit{(1) State space representation.} 
Most blockchains are structured as a directed acyclic graph (DAG), so a naive state space representation might be to use the entire DAG as the current state.
This approach has a few problems.
First, its dimensionality grows over time.
Second, it includes irrelevant information (e.g., old side chains that cannot be displaced with high probability \cite{backbone}). 
We therefore require a representation of the state space that is general enough to apply to different blockchains, while  being constrained enough to limit the problem dimension.
An important part of \name{} is our derivation of a general framework for extracting a compact state space representation (features) that is general enough to learn meaningful attacks for different protocols. 

\textit{(2) Learning algorithm.} 
Two common classes of learning algorithms are value-based methods and policy gradient methods \protect\cite{franccois2018introduction}.
Value-based methods typically aim to build a \emph{value function} that associates some value with each state; a common example is Q-learning \protect\cite{q-learning}.
Policy gradient methods instead try to optimize rewards by performing gradient ascent on a parametric policy, which  in our case will be represented by a neural network \cite{franccois2018introduction}.
Common examples include the REINFORCE algorithm \protect\cite{williams1992simple} and actor-critic methods \cite{sutton2000policy}.

We  make  use of both classes of algorithms in this work, as
different DRL algorithms perform well in different settings. 
The most basic algorithm is deep Q-networks (DQNs), which is based on the classical idea of Q-learning \cite{q-learning}.
However, more sophisticated algorithms (including policy gradient methods) have surpassed DQNs in many problem areas \cite{ppo, acktr, a2c}, sometimes at the expense of higher computational cost. 

\textit{(3) Reward function.} 
Designing a good reward function can significantly impact the success of the overall system; often this requires a combination of domain knowledge and some tuning of hyperparameters.


\textit{(4) Training heuristics and hyperparameters.}
Some attack models and/or protocols can be difficult to learn due to complexity.  Leveraging blockchain domain knowledge to design training heuristics can help DRL agents learn, which we show in \Cref{ssec:model}.


\section{\name: System description}
\label{sec:system}

\begin{figure}
    \centering
    \includegraphics[width=\columnwidth]{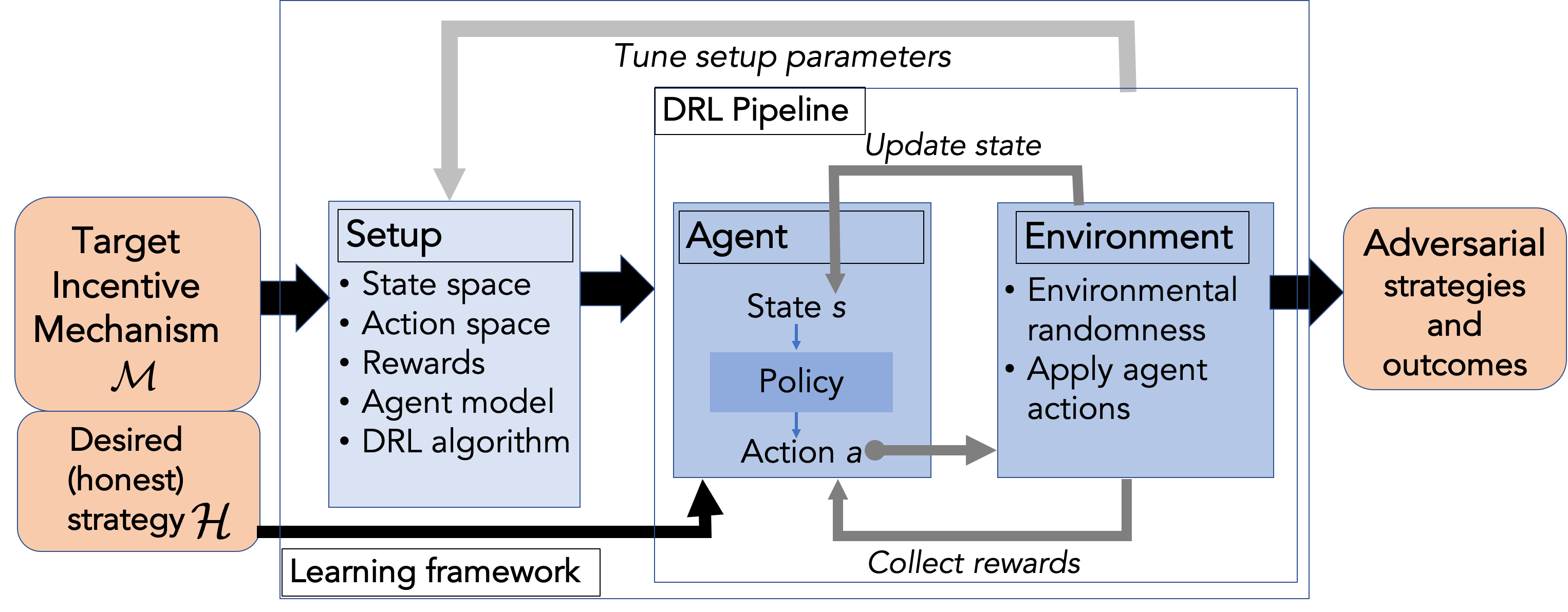}
    \caption{Schematic of \name~learning framework. 
}
    \label{fig:framework}
\end{figure}



Figure~\ref{fig:framework} shows use of the RL-based {\em learning framework} that is the cornerstone of \name. It involves a three-stage pipeline for discovering and analyzing adversarial strategies targeting an incentive mechanism $\mechanism$. 

First, the protocol designer builds an {\em environment} that simulates execution of the protocol realizing $\mechanism$. We anticipate that the bulk of the effort involved in the \name~framework will  go into this part of the system, as the environment fully encapsulates a model of \mechanism. The protocol designer instantiates in the environment a set of {\em features} (state space respresentation) over which learning will occur, as well as a space of {\em actions} that agents may take. 
(What features to use may be iteratively tuned to improve performance of the RL algorithm at later stage.) 
In parallel, the protocol designer chooses an {\em adversarial model} to explore. We have outlined a principled set of choices in our discussion of  Table~\ref{table:experimental_progression} in Section \ref{sec:motivation}. 
Finally, the protocol designer selects an {\em RL algorithm}  appropriate for the environment and adversarial model. She must associate with the RL algorithm a reward function and hyperparameters, both of which may be iteratively tuned as exploration proceeds. 

In this section, we describe respectively the environments, adversarial models, and RL algorithms we employ in  \name{} for the experiments in this paper. 
%
We focus on incentive mechanisms related to block rewards since these have dominated the literature on  blockchain incentive mechanism vulnerabilities \cite{eyal2018majority,sapirshtein2016optimal,nayak2016stubborn,grunspan2019selfish,semiselfish,eyal2015miner}, but similar ideas can be applied to transaction rewards, for example.
The environment construction depends on the precise incentive mechanism (and underlying consensus algorithm), but there is a strong commonality among our environment designs. 
In particular, we describe four aspects of the environment:  
(1) the block generation model, (2) the reward function, (3) the action space, and (4) feature extraction. 
Of these, the most challenging design problem is the feature extraction, 
so we provide a general-purpose algorithm for choosing a feature extractor given an action space and a model of the environmental randomness. 

\subsection{Blockchain generation}
As in prior work \cite{eyal2018majority,sapirshtein2016optimal}, we assume a randomized model for block generation. 
Hence, we view block generation as a discrete-time process, where a new block is generated at each time slot $i\geq 0$ (we generalize this assumption in \Cref{sec:rushing}).
In proof-of-work (PoW), a party that controls fraction $\alpha$ of the network's mining  power mines the $i$th block with probability $\alpha$, independently across all time slots.
This model is easily extended to PoS, where parameters depend on the block generation mechanism.

In accordance with prior work, we assume that the network communicates blocks instantaneously to other nodes (we relax this assumption in \Cref{sec:casper}).
In the event that an agent's block is received simultaneously with an honest party's block, and both are considered equally viable by the consensus mechanism $\mechanism$, we assume the honest nodes all follow the adversarial block with probability $\gamma$, and all follow the honest block with probability $1-\gamma$.
We call $\gamma$ the \emph{follower fraction}.

\subsection{Rewards} 
In most chain-based blockchains (e.g., Bitcoin), the miner of a block that appears in the final ledger receives a \emph{block reward}. 
Like prior work  \cite{eyal2018majority,sapirshtein2016optimal,nayak2016stubborn}, we consider one or more attackers that aim to maximize their block rewards given a constrained amount of computational resources; we ignore  transaction fees for simplicity. 
We define $B_a(t, \stratagent)$ and $B_o(t, \stratagent)$ as the rewards of the attacker of interest and all the other miners, respectively, at time $t$ under an attack strategy \stratagent.
When there are multiple attackers of interest, we differentiate them with superscripts.
We compute a miner's reward by aggregating the block reward it accumulates, avoiding analysis of uncontrolled externalities such as coin price or electricity cost.  

In practice, miners appear to optimize their \textbf{absolute reward rate}, defined as $\lim_{t\to \infty} \frac{B_a(t,\stratagent)}{t}$. However, prior work \cite{eyal2018majority, sapirshtein2016optimal} has mostly focused on \textbf{relative rewards}, defined as the attacker's block rewards as  a fraction of the whole network's rewards: $\lim_{t\to \infty} \frac{B_a(t,\stratagent)}{B_a(t,\stratagent)+B_o(t,\stratagent)}$. 
In \Cref{app:near}, we show that over even moderate time periods in the Bitcoin protocol, the two objectives are interchangeable. 
We will focus on relative rewards for many of our results, since this enables direct comparison with prior work.
\vspace{1mm}





\subsection{Action space and adversarial model}
There is a close tie between the action space and adversarial model. 
DRL requires system designers to specify the action space,
which is informed in part by the type of attacks one is searching for. 
Typically, in block-reward-based incentive mechanisms, we expect an agent to be able to control, at a minimum, where to append blocks on its local view of the blockchain, and when to release them.

\smallskip
\noindent \textit{Example (Bitcoin):}
We allow agents to take one of four basic actions, as  in \cite{eyal2018majority,sapirshtein2016optimal}.
The first is \textit{adopt}, in which an agent abandons its private fork to mine on the canonical public chain.  This action is always allowed.  
The second action is \textit{override}, in which an agent publishes just enough blocks from its private chain to overtake the canonical public chain.  
This action is feasible only if the agent's private chain is strictly longer than the public main chain.
The third action is \textit{wait}, in which an agent continues to mine without publishing any blocks.  This action is always feasible.  
The  final action is \textit{match}, in which an agent publishes just enough blocks to equal the length of the longest public chain as the block on the longest public chain is being published, causing a fork.  In \cite{sapirshtein2016optimal}, the authors prove the optimality of this action space in single-strategic-agent selfish mining  (i.e., if an action outside of this action space is chosen, it results in strictly lower rewards).

\vspace{0.1in}
In addition to specifying the action space, the adversarial model requires system designers to specify the number of strategic agents and their relative resources.
In our experiments on selfish mining in Sections~\ref{sec:eval} and~\ref{ssec:eval_multi}, we explore the numbers and types of adversaries given in Table~\ref{table:experimental_progression}, for example. 

The type of each agent $i$ in our various experiments is specified in part by a  (fractional) hash power $\alpha_i$. 
In our selfish mining experiments, agent types also include  an agent-specific ``follower fraction'' $\gamma_i$ that specifies the probability that honest nodes follow a particular agent's chain in the case of multi-way ties; 
this parameter models an agent's network penetration and generalizes the previously-described parameter $\gamma$.


\subsection{State space representation: Feature extraction}
\label{sec:state-space}
The goal of \name~is to find a policy (or  strategy) $\pi: S \to \Delta(A)$,  where $\Delta(A)$ denotes the probability simplex over actions in $A$. We let $\pi(\cdot | s)$ denote the mapping from a state $s$ to a distribution over actions.
Ideally, we would implement this $\pi$ as a lookup table. 
However, in general, the state space $S$ is the set of all possible blockDAGs, which is infinite. 

The canonical way of solving this problem is with function approximation \cite{csababook}; instead of making $\pi$ a lookup table, we implement it with a different function.  Concretely, for some $d\in \mathbb N$, let $f: \mathbb{R}^d \to \Delta(A)$, and $\varphi: S \to \mathbb{R}^d$.  We then let $\pi(\cdot | s) = f(\varphi(s))$.  $f$ is usually chosen to be a neural network.  $\varphi$ is called the feature extraction method, and designing a good $\varphi$ for blockchains will be the focus of this section.

We write a general framework for specifying $\varphi$ in blockchain incentive mechanism problems.
Our suggestions for how to construct $\varphi$ may be lossy or redundant in general; however, we will show that for a variety of blockchain incentive problems, it can be used by \name{} to achieve near-optimal performance.

To motivate our procedure, 
we want to design features that are descriptive enough to find attacks on protocols. 
Recall that in an MDP, the value function $V: S \to \mathbb R$ is defined as 
\begin{align*}
    V(s) := \E[\sum_{t = 0}^\infty \eta^t R_{a_t^*} (s_t, s_{t + 1})|s_0 = s]
\end{align*}
where $a_t^*$ is the optimal action at time step $t$ wrt the entire process starting from time 0 \cite{csababook}.    
We want $\varphi$ such that 
\begin{align*}
    \inf_g \E_{s \sim \mu}[(V(s) - g(\varphi(s)))^2] < \epsilon
\end{align*}
for some  small $\epsilon \geq 0$, where $\mu$ is some distribution over initial states.  
Define $W_\varphi(s) := \{s': \varphi(s') = \varphi(s)\}$.  If $g(\varphi(s')) = V(s)$ for all $s' \in W(s)$, then it is clear that $\epsilon = 0$.  This suggests a procedure for choosing $\varphi$: choose it so that states $s$ that map to the same value of $\varphi(s)$ have the same $V(s)$ (or close to the same value).  In other words, we want to find the features of $s$, $\varphi(s)$ that determine $V(s)$ relatively closely.  Then our optimal value approximator $\hat{g}$ would simply be $V \varphi^{-1}$, where the inverse maps to any state in the pre-image of $\varphi$.  

Once we define a suitable $\varphi$, we can either rely on DQNs to learn  $\hat g$ directly (which in turn gives a policy), or we can use policy gradient methods to directly learn a strategy that operates on the features $\varphi(s)$.  

We assume that some subset of blocks in each state $s$ belongs to the agent. In most blockchains, the value of the state $V(s)$ is determined by a few properties:  
(1) \emph{score}: how likely is a protocol to choose the agent's subset as the canonical one? 
(2) \emph{instantaneous reward}: how much reward would the agent receive if its subset is chosen?
(3) \emph{permitted actions}: what actions are allowed in a given state?  
Note that to estimate the value function, an agent should track these quantities both for its own blocks and for the visible blocks of the other players. 


\subsubsection{Score}
Given a blockDAG $T$,  the score of connected subgraph  $C \subseteq T$ determines $C$'s chance of being selected as the canonical subgraph, i.e., the subgraph on which honest nodes build.  
The score of subgraph $C$ can also depend on observable environmental variables  $E=(e_1,\ldots,e_m)$. 
For example, these  could represent an agent's level of resource allocation.
Hence, we define \emph{score} function $L(C,T,E)$ as a mapping from $(C,T,E)$ to a $r$-dimensional vector.  
We propose to let $\varphi(s)[1: r] = L(C,T,E)$, where we use array notation to show that the first $r$ dimensions of the features are the score.

\smallskip

\noindent \textit{Example (Bitcoin, Ethereum, Fruitchains):}  In these protocols, the fork with the longest length is adopted as the canonical chain.  So $L(C,T,E) = [\textrm{len}(C)]$ where $\textrm{len}(C)$ is the length from the last globally accepted block and the block that the agent is currently mining on.

\smallskip

\noindent \textit{Example (GHOST \cite{ghost}):} GHOST chooses the heaviest subtree at each node to arrive at a canonical chain.  Let $G$ be the set of nodes that GHOST would traverse by choosing the heaviest weighted subtree at each node, assuming that all blocks were known to GHOST, including any of the agent's unpublished blocks.  Then we can let $L(C,T,E) = [|G \cap C|]$.

\smallskip \textit{Example (Casper FFG + Ethereum):}  Two things impact the likelihood of a chain becoming the canonical chain in this protocol: its length, and the proportion of the votes that it has received.  So $L(C,T,E) = [\textrm{len}(C), v]$ where $v$ is the proportion of votes that the agent has received so far for a checkpoint on its chain.

\begin{figure}
    \centering
    \includegraphics[width=1.5in]{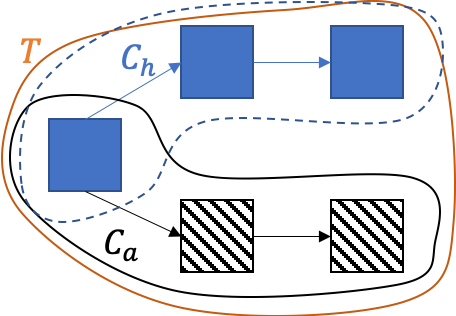}
    \caption{Sample state for Bitcoin. Subchain $C_h$ (blue solid blocks) denotes the public (honest) main chain, and subchain $C_a$ (black striped blocks) denotes the agent's private chain. }
    \label{fig:state_btc}
\end{figure}

\smallskip

\subsubsection{Instantaneous reward}
Blockchain protocols vary in block reward allocation schemes, ranging from rewarding only longest-chain blocks (e.g., Bitcoin, GHOST), to uncle-based rewards (e.g., Ethereum \cite{wood2014ethereum}, Fruitchains \cite{pass2017fruitchains}).  These reward mechanisms directly impact the values of a state, so they must also be part of feature extraction.

One commonality among block reward schemes is that they are awarded on a per-unit basis for some unit (blocks, fruits, uncles, etc.).  
Define $U(s)$ as the set of units that are relevant to the agent in the current state.  Then define the features of these units (e.g. how recently they were mined, their reward content, etc) as $\textrm{feat}(U(s))$, where the dimension of $\textrm{feat}(U(s))$ is $r' \in \mathbb N$.  We propose to include these features in our feature mapping $\varphi$.  Concretely, we let $\varphi(s)[1] = L(C,T,E)$ and $\varphi(s)[r + 1:r + r' + 1] = \textrm{feat}(U(s))$. 

In general, this can give a high-dimensional feature mapping, which will cause training to be hard.  However, in many protocols, $\textrm{feat}(U(s))$ can be very low dimensional, as we demonstrate in the following example.

\smallskip

\noindent \textit{Example (Bitcoin):} Block rewards in Bitcoin are awarded equally to every block on the canonical chain.  Therefore, for a selfish miner in the single-agent setting, $\text{feat}(U(s)) = [\sum_{B \in C} \mathbbm{1}\{B \notin M\}] = [\textrm{len(C)}]$, where $M$ is the established canonical chain and $B$ is a block.  

\subsubsection{Permitted actions}
From a given state $s$, an agent may only be allowed to take a subset of actions in $A$, which influences $s$'s value.  For instance, in  \cite{sapirshtein2016optimal}, an agent can only match (i.e. cause a tie between two longest chains) if it already had an unpublished block at the same height as a block the honest party publishes.

We define permitted actions as a binary array $\textrm{act}(s) \in \{0,1\}^{|A|}$ where $|A|$ is the size of the action space.  $\textrm{act}(s)[i] = 1$ if the $ith$ action is available at state $s$ and $0$ otherwise. We propose to have $\varphi(s)[r + r'+2:r + r'+2 + |A|] = \textrm{act}(s)$.
While $|A|$ is usually small, it can sometimes be compressed.  

\smallskip
\noindent \textit{Example (Bitcoin):}
With the score features from both the honest party and the strategic agent, as well as observed environmental variables, we can completely determine the permitted actions of a state.  Define an environmental variable $fork \in \{{\tt relevant}, {\tt irrelevant}, {\tt active}\}$, which 
describes the most recent event in the system \cite{eyal2018majority,sapirshtein2016optimal}.
$fork= {\tt relevant }$ if an honest node just mined a block.
 $fork= {\tt active }$ if the agent just tried to match another party's block;
 in this case, recall that the environment chooses the agent's block over the honest one w.p. $\gamma$.  
Otherwise,   $fork= {\tt irrelevant }$.   \cite{sapirshtein2016optimal} showed that these three features are sufficient for determining the permitted actions of a state.  Because we are already storing scores, we only need one additional feature, $fork$, in place of the permitted actions array.  

\smallskip
The described Bitcoin feature mapping matches the feature mapping in  \cite{sapirshtein2016optimal} and was shown in \cite{sapirshtein2016optimal} to achieve $\epsilon = 0$.  
In other words, policy iteration was able to find a $g$ such that $g(\varphi(s)) = V(s)$ for all $s$.  Therefore, in this case, our procedure found a sufficiently descriptive feature mapping.  We also provide an instantiation for Ethereum in \Cref{sec:ethereum}, and demonstrate the features' sufficiency empirically by surpassing state-of-the-art selfish mining rewards for Ethereum.

\subsection{Order of operations}

An important adjunct to design of the action space and feature extraction is a policy for  {\em sequencing} agents' actions and disseminating rewards. 
In our environment, we enforce three properties that are motivated by our problem domain: 

\noindent \textit{(1) Synchronous action selection.} We assume all actions of strategic agents are recorded synchronously, after seeing the actions of the honest party (if it exists).  
This is needed to prevent the adversary from observing the actions of other strategic players and reacting accordingly; in this case, the last strategic agent to choose its action would have an unfair advantage.
However, we do allow for a rushing adversary who sees the honest party's actions before deciding how to act, consistent with \cite{sapirshtein2016optimal,semiselfish}.
In \Cref{sec:rushing}, we illustrate how this assumption breaks down in the multi-agent setting.

\noindent \textit{(2) Delayed execution of passive behavior.} 
Once actions are recorded, we must apply them in some order. 
We have chosen  always to apply the "passive"  action last.  
To see why, consider the following scenario:
In Bitcoin, suppose an honest agent is mining on the public chain in the presence of two strategic players, Alice and Bob. 
Now suppose Alice overrides and Bob adopts (i.e., takes the passive action).  
We know that actions are collected synchronously, but if they were also \emph{processed} synchronously, Alice's chain would become the main chain, while Bob would have adopted the previous main chain, which is now stale.
This behavior is unrealistic because a strategic Bob would choose to mine on Alice's override block if it were to abandon its private chain.

\noindent \textit{(3) Delayed multi-agent rewards.} 
In the multi-agent setting, rewards should not be immediately allocated following an override or adopt action.  
Unlike the single-agent setting, there is still a possibility of the strategic party's chain being overridden by another strategic party.
Hence, we only allocate a block reward if all agents acknowledge the block, i.e., adopt it.

\subsection{RL algorithm}
\label{sec:rl_algo}

We employ different deep reinforcement learning algorithms, depending on the adversarial model. In the single agent setting, we use Deep Dueling Q-Networks (DDQN) \cite{ddqn} and in the multiple agent setting, we use Proximal Policy Optimization (PPO) \cite{ppo}.  
In our experiments, we have found that DDQN converges faster than PPO in the single agent setting for Bitcoin: with a block limit of 5, $\alpha = 0.4$, $\gamma = 0$, DDQN converges in roughly $10^{5}$ steps in the environment, while PPO takes an order of magnitude more steps to converge.

However, DDQN can fail when there are multiple adaptive agents because the Markov assumption no longer holds.
Although PPO is not immune to this problem, it has been used successfully for multi-agent games \cite{dota,berner2019dota,baker2019emergent}, and we found it to be more stable than the alternatives in the multi-agent setting.

\subsection{Implementation}
We used OpenAI Gym \cite{gym} to construct our environments and execute RL algorithms on them.   
OpenAI Gym provides a generic interface for implementing environments. 
In our case, this environment specifies a model for the target incentive mechanism \mechanism. 
The environments we have implemented provide a template for users to easily instantiate their own blockchain protocols (namely, blockchain structures ranging from pure chains to generic DAGs).
We use the RLLib~\cite{rllib} training interface to train our agents on state-of-the-art RL algorithms and list the relevant hyperparameters for the following experiments.

\begin{figure}[t]
\centering
\includegraphics[width=0.8\columnwidth]{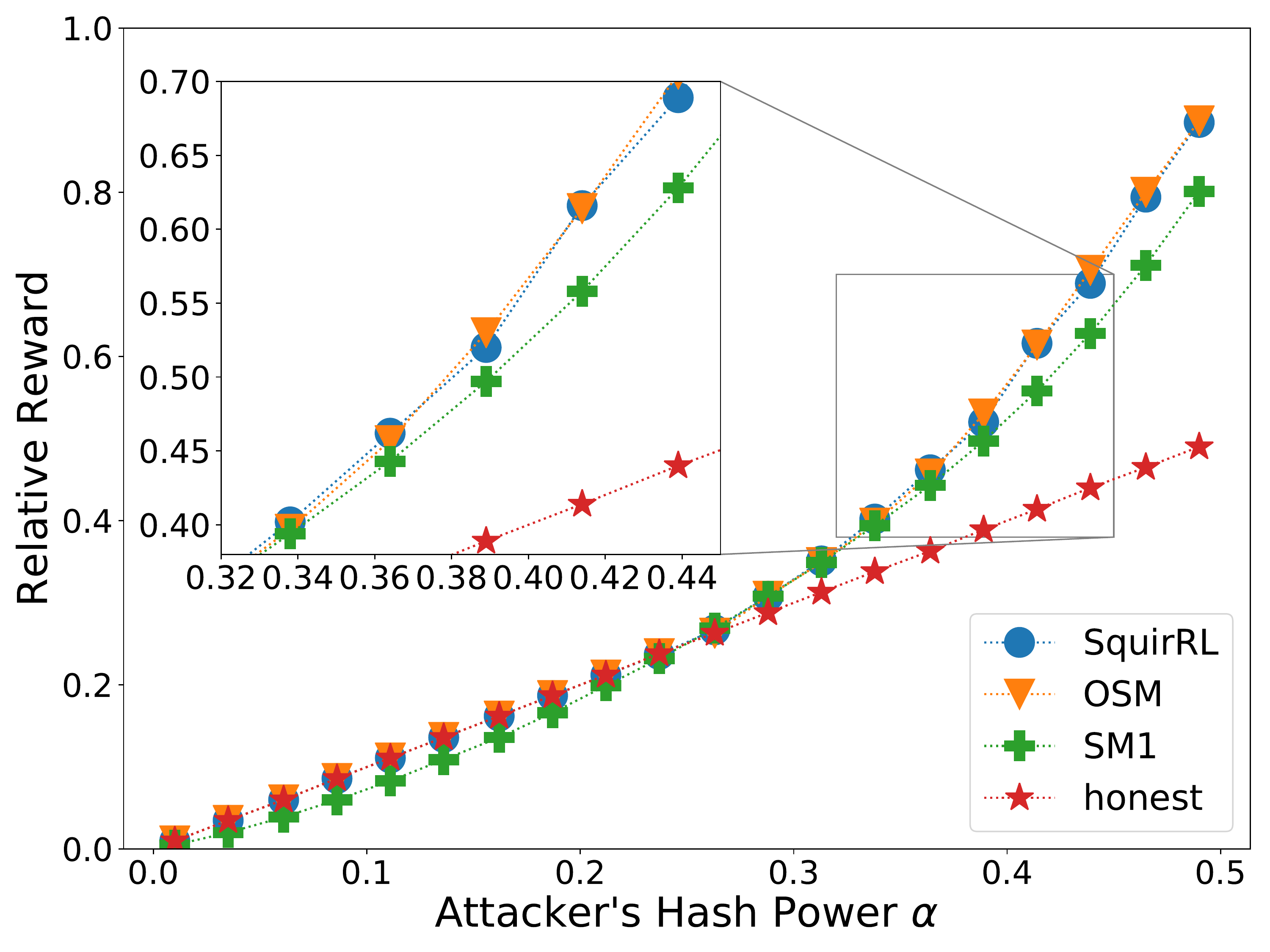}
\caption{Bitcoin relative reward as a function of  adversarial hash power. \name~recovers the findings of \protect{\cite{sapirshtein2016optimal}}.}
    \label{fig:btc_r_vs_alpha}
\end{figure}

\begin{figure}[t]
\centering
\includegraphics[width=0.8\columnwidth]{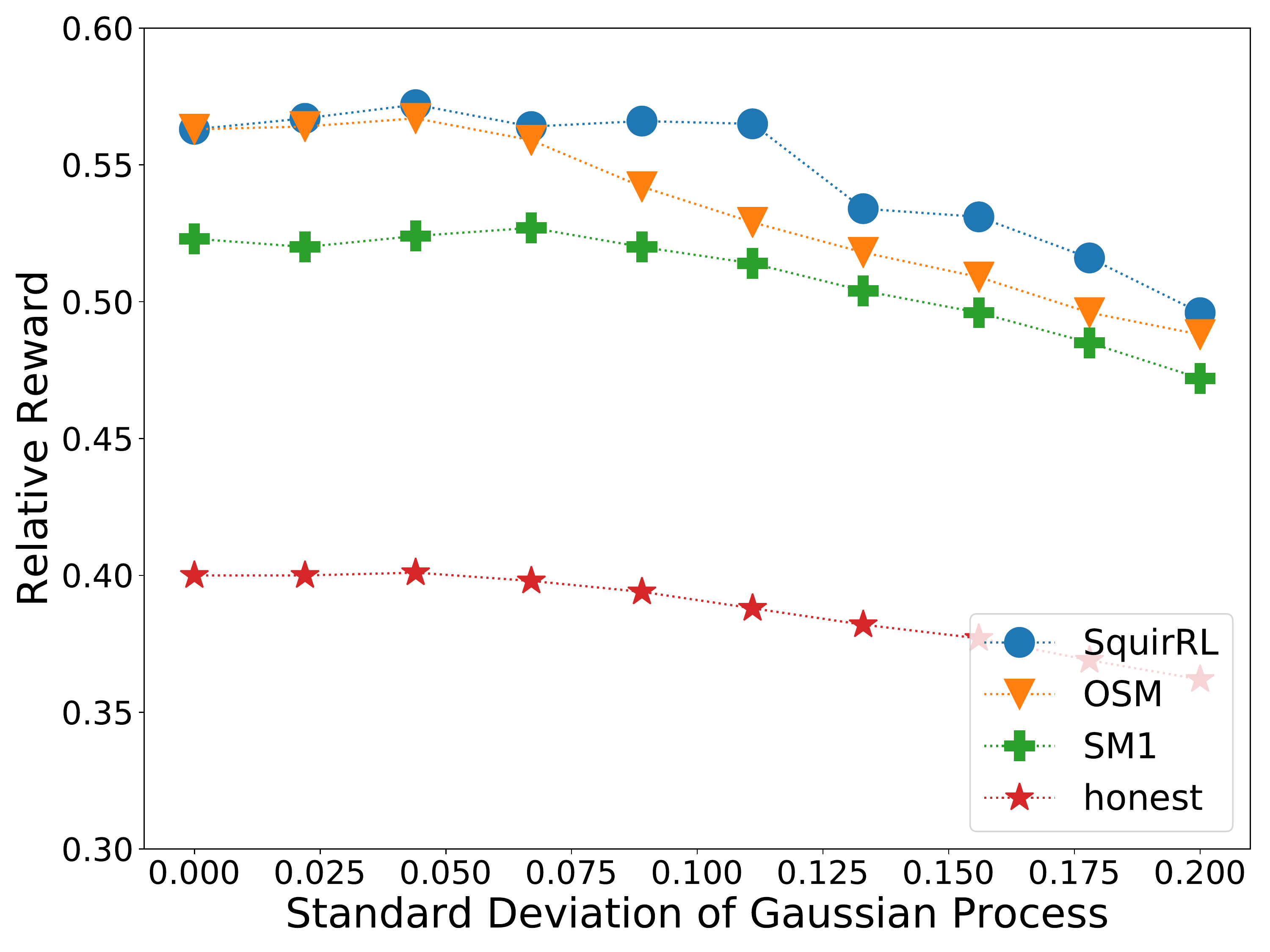}
\caption{Bitcoin relative reward under stochastic $\alpha$ (Gaussian random process with $\mathbb{E}[\alpha]=0.4$). 
}
    \label{fig:btc_stoch_alpha}
\end{figure}

\section{Evaluation: Single Strategic Agent}
\label{sec:eval}
We first consider applications of \name~to selfish mining attacks when there is a single strategic agent, and the remaining agent(s) follow protocol \protect{\cite{eyal2018majority,sapirshtein2016optimal,nayak2016stubborn}}.
We will focus here on the Bitcoin protocol; analogous experiments on Ethereum can be found in \Cref{sec:ethereum}.
In our first batch of experiments, the main benefit of DRL over algorithms for solving MDPs (e.g.,  value iteration) is that  DRL can handle larger  state spaces.
We then demonstrate that DRL learns good strategies in a stochastically varying environment of unknown distribution, which is not possible with an MDP.
These experiments lay the groundwork for Section~\ref{ssec:eval_multi}, where we describe experiments with multiple strategic selfish mining parties.
Our experiments compare to several baseline mining strategies: 

%

\noindent \textbf{(1) Honest mining:} a miner who follows protocol. 

\noindent \textbf{(2) Optimal selfish mining  (OSM):} the strategy learned in \protect{\cite{sapirshtein2016optimal}} for the Bitcoin  protocol. 

\noindent \textbf{(3) SM1:} the selfish mining strategy originally proposed in \protect{\cite{eyal2018majority}}; although this baseline should be strictly dominated by  OSM in the Bitcoin  setting, it has been used in  other settings as well \protect{\cite{niu2019selfish}};  we include it for completeness. 

\noindent \textbf{(4) \name:} the strategies output by our training pipeline. 


In each of our experiments, we train a DRL agent and simulate all baseline strategies to compete. 
Then, for a given parameter setting (e.g., initial adversarial  party's fraction of  hash power $\alpha$), we run 100 trials of each blockchain protocol, where each trial consists of 10000 
steps in the MDP and generates at least 5000 blocks in the main chain. 
We compute the relevant parties' rewards for each trial, and average over all trials. 

\subsection{Static Hash Power}
\label{sec:eval_bitcoin}
The Bitcoin  protocol is a useful case study in  part because its incentive mechanism is well-studied \protect{\cite{eyal2018majority,sapirshtein2016optimal,nayak2016stubborn,eyal2015miner}}. 
Prior work has recovered an \emph{optimal} selfish mining strategy in the one-strategic-agent case when hash power is static \protect{\cite{sapirshtein2016optimal}}.
A useful sanity check is thus to see if \name~recovers these known optimal results. 
In~\protect{\cite{sapirshtein2016optimal}}, the authors recover the optimal strategy for selfish mining in Bitcoin by casting the problem as an MDP  and applying policy iteration.
We aim to recover and replicate two key findings of their work: 

\noindent \textit{(1)} Selfish mining is only profitable for adversaries who hold at least $25\%$ of the stake in the system; this assumes that if the adversary publishes a block at the same time and height as the honest chain, the honest nodes will  build on the adversary's block with probability $\gamma=0.5$. An adversary who holds less than  $25\%$ of the stake should revert to honest mining.

\noindent \textit{(2)} For adversaries with  more than $25\%$ of the stake, the authors of~\protect{\cite{sapirshtein2016optimal}} show performance curves that quantify the adversary's relative increase in rewards compared to honest mining. 
    Our  goal  is to match these curves.

\smallskip
Figure \ref{fig:btc_r_vs_alpha} demonstrates the outcome of this experiment, using the state space developed in \cref{sec:state-space}. 
We observe two key findings.  
First, for $\alpha < 0.25$, \name~does not adopt a 'selfish mining' strategy, but recovers the honest mining strategy.
This sanity check is consistent with the theoretical findings of \protect{\cite{sapirshtein2016optimal}}.
Second, for $\alpha>  0.25$, we find that \name~achieves a relative reward within 1\% of the true optimal mechanism. 
This result required minimal tuning of hyperparameters.

We find similar results for the Ethereum blockchain in \Cref{sec:ethereum}. 
Ethereum's larger state space (more complex reward mechanism) makes it poorly-suited to value iteration. Here, \name{} easily recovers strategies with higher rewards than state-of-the-art approaches \cite{ritz2018impact,niu2019selfish,grunspan2019selfish}.

\begin{table}[t]
\centering
\resizebox{\columnwidth}{!}{%
\begin{tabular}{@{}lrrrr@{}}
    & \multicolumn{4}{c}{{\bf Strategy}} \\
    \cmidrule{2-5}
    {\bf Name} & Honest & SM1   & OSM   & \name{}    \\ 
    \toprule
    Bitcoin  & 0.398 $\pm$ 0.008 & 0.540 $\pm$ 0.016 & 0.566 $\pm$ 0.018 & \textbf{0.585 $\pm$ 0.019} \\ 
    Monacoin & 0.407 $\pm$ 0.007 & 0.552 $\pm$ 0.017 & 0.594 $\pm$ 0.020& \textbf{0.602 $\pm$ 0.020} \\ 
    Vertcoin & 0.406 $\pm$ 0.007  & 0.554 $\pm$ 0.014 & 0.597 $\pm$ 0.020 & \textbf{0.602 $\pm$ 0.020}  \\ 
    Litecoin & 0.408 $\pm$ 0.007 & 0.564 $\pm$ 0.016 & 0.603 $\pm$ 0.019 & \textbf{0.608 $\pm$ 0.022}  \\
    \bottomrule
    \end{tabular}
}
\caption{Relative rewards under stochastic $\alpha$ as measured in real cryptocurrencies from September 24-October 28, 2019. Results shown for initial $\alpha=0.4$. We show the average and the standard deviation results for 100 repetitions.}

\label{tab:real_data}
\end{table}

\subsection{Variable Hash Power}
\label{sec:stoch_alpha}

In the previous experiments (Bitcoin, Ethereum), it is possible to write an MDP approximating the system dynamics (even if the state space is large).  
In more realistic blockchain settings, the underlying MDP may be changing over time or unknown.  
In this section we explore such a scenario, where the adversary's hash power $\alpha$  \emph{changes} stochastically over time. 
This can happen, for instance, if the adversary maintains a fixed amount of hash power (in megahashes/day) while the total hash power in the cryptocurrency fluctuates, or if miners dynamically re-allocate hash power over time to different blockchains \protect{\cite{kiraly2018profitability, meshkov2017short}}. 
In either scenario, formulating an MDP is challenging for two reasons: (1) We may not know the distribution of random process $\alpha(t)$; 
(2) Even if we can estimate it (e.g. from historical data), incorporating this continuous random process into an MDP would bloat the feature space.

\name~handles this uncertainty 
by using the current value of $\alpha$ during training without any knowledge of the underlying random process. 
We find that \name~learns more robust strategies than those in the literature and is therefore less likely to overreact to outlying values. 

We evaluate performance for stochastic $\alpha$ by first allowing $\alpha(t)$ to vary according to a Gaussian white noise random process with  $\mathbb E[\alpha(t)]=0.4$. 
Figure \ref{fig:btc_stoch_alpha} illustrates the relative reward as a function of the standard deviation of this process. 
We truncate fluctuations to $\alpha\leq 0.5$ to avoid 51\% attacks. 
When $\alpha(t)$ has low variance, our results are consistent with those from  Section~\ref{sec:eval_bitcoin}: \name~achieves relative rewards close but not identical to OSM. 
However, as the variance increases, \name~actually starts to outperform OSM.
Intuitively, the learned strategies are less likely  to react to fluctuations in $\alpha(t)$, thereby preventing the agent from taking extreme actions for anomalous events.
We would consequently expect \name~to perform particularly well on blockchains with low (and hence more volatile) total hash power. 

To explore the effect of stochastic $\alpha$ in the wild, we ran \name~on data from real cryptocurrencies. We first trained \name~in an environment where $\alpha(t)$ followed a Gaussian white noise random process with standard deviation 0.1, a value chosen
prior to seeing the data  from real cryptocurrencies.
Next, we scraped the estimated total hash power hourly for at least a month for four blockchains that use Bitcoin's consensus protocol and block reward mechanism:
Bitcoin, Litecoin, Monacoin, and Vertcoin. (See~\cref{app:hash_power} for detailed data.)
We then assumed an attacker with constant raw hash power (in MH/day); this raw hash power is chosen by initializing the attacker at a relative hash power of $\alpha=0.4$ in each measured blockchain. 
Once the absolute hash power is fixed, the attacker's relative hash power $\alpha$ fluctuates  solely due to changes in the total hash power of each blockchain. 
Table \ref{tab:real_data} shows the relative rewards resulting from various strategies.
\name~achieves the highest relative rewards (although within statistical error of OSM), showing RL's benefits in environments that change in ways difficult to capture with an MDP. 



\section{Multi-Agent Selfish Mining Evaluation}
\label{ssec:eval_multi}


The previous section demonstrated the ability of \name{} to (a) learn a known optimal strategy for Bitcoin, (b) extend prior state-of-the-art results on Ethereum in a setting where the state space is too large for an MDP solver, and (c) learn strategies in a stochastic, possibly nonstationary environment. 

In this section, we instead demonstrate DRL's ability to handle nonstationary environments in which \emph{multiple} strategic agents are competing in the Bitcoin selfish mining scenario. 
This section has three main findings. In a multi-strategic-agent setting: (1) OSM is not a Nash equilibrium. (2) The commonly-studied \emph{rushing adversary} can have counterintuitive and nonphysical results. This has general implications for how the research community should model multi-agent security problems moving forward. 
(3) We do not observe any benefit to selfish mining when $k\geq 3$ strategic agents are competing. This suggests that even over an infinite time horizon, selfish mining is not a serious attack for the Bitcoin protocol.
\subsection{Model}
\label{ssec:model}

We generalize the model from Section \ref{sec:system}.
Recall that for a single strategic agent, we used $\gamma$ to denote the probability of the honest party choosing an adversarial block over an honest one in the event of a match. 
For the multi-agent setting, we instead define the \emph{follower fraction} $\gamma_i$, which we briefly described above. For $i\in \{1,\ldots, k\}$, $\gamma_i$ is the probability of the honest agent building on the $i$th agent's block in  case of a $k$-way tie. 
This models each party's network connectivity.
In case of a tie among fewer parties, the $\gamma_i$ values are  normalized appropriately.

The multi-agent setting requires a different abstraction than MDPs: Partially Observed Markov Game (POMG)~\cite{zhang2019multi}.  
A POMG is a tuple $(N, S, \{A_i\}_{1 \leq i \leq N}, P, \{R^i\}_{1 \leq i \leq N}, \Omega, \{O^i\}_{1 \leq i \leq N})$, where $N$ denotes the number of agents, $S$ is the state space for all the agents, $A_i$ is the action space for agent $i$, $P: S \times A_1 \times \cdots \times A_N \times S \to \mathbb R$ denotes the transition probability from a state $s \in S$ to $s' \in S$ given joint action $a \in A_1 \times \cdots \times A_N$, $R^i: S \times A_1 \times \cdots \times A_N  \times S \to \mathbb R$ is the reward function for agent $i$ that determines the immediate reward transitioning from state $s \in S$ to state $s' \in S$ with joint action $a \in A_1 \times \cdots \times A_N$, $\Omega$ is the space of observations, and $O^i: S \to \Omega$ maps the state to the observation that agent $i$ sees.

In our setting, $S$ is the space of all possible blocktrees.  $A_i = A$ for all $i$, where $A$ was the action space from the single-strategic-agent setting.  $R^i(s, a, s') = (1 - \alpha_i)x_i - \alpha_i y_i$ where $x_i$ is the number of blocks agent $i$ received in the process of transitioning from $s \in S$ to $s' \in S$ with joint action $a \in A_1 \times \cdots \times A_N$.  $\Omega$ is the space of blocktrees except without the hidden chains included.  $O^i(s)$ is the blocktree observed by agent $i$ (note that we use feature extraction as before, except applied to the observations, to simplify this representation).

But what is $N$?  It should include all of the strategic agents, of course, but should it include the honest party?  In \cite{eyal2018majority,sapirshtein2016optimal} (which study the single-agent setting) the honest party is treated as part of the environment and accounted for in the probability transition matrix and in the reward matrix.  In recent work on multi-agent selfish mining \cite{semiselfish}, the honest party is also considered to be a part of the environment.  

Consider a POMG with honest party $A$ and strategic agents $B$ and $C$.  If the honest party is in the environment, then upon receiving the joint observation $o \in \Omega^2$, agents $B$ and $C$ submit the joint action $a \in A^2$.  After the actions are submitted, a block is awarded to one of $A,B,C$ and $A$ performs an action (notably, before $B$ and $C$ perform any more actions or observe anything).  Rewards are given to the agents.  Then agents receive another observation $o \in \Omega^2$, and the cycle repeats.
Notice that even in the DRL literature, it is standard to represent agents with known strategies as part of the environment \cite{dota,littman1994markov}.

If $A$ is outside of the environment (i.e., included as an agent), then upon receiving the joint observation $o \in \Omega^3$, agents $A,B,C$ submit joint action $a \in A^3$.  After actions are submitted, a block is awarded to one of the parties and  rewards are allocated.  Then the cycle repeats.  The difference is in when the strategic agents can see the hidden state of $A$: when $A$ was part of the environment, $A$ never had a hidden chain because it publishes while the POMG processes the transition.  When $A$ is considered an agent, it doesn't publish its block until the next turn, which means its hidden chain length could be nonzero in length.  

As we are interested in studying worst-case security, it is tempting to think that giving $B,C$ more information is a more conservative choice, thus implementing the honest party as part of the environment. 
We start with this assumption to be consistent with \cite{sapirshtein2016optimal,semiselfish}.
However, we show in \Cref{sec:rushing} that this decision can lead to counterintuitive results. 

\begin{figure*}[t]
	\begin{minipage}[b]{0.33\linewidth}
		\centering
		\includegraphics[width=0.9\textwidth]{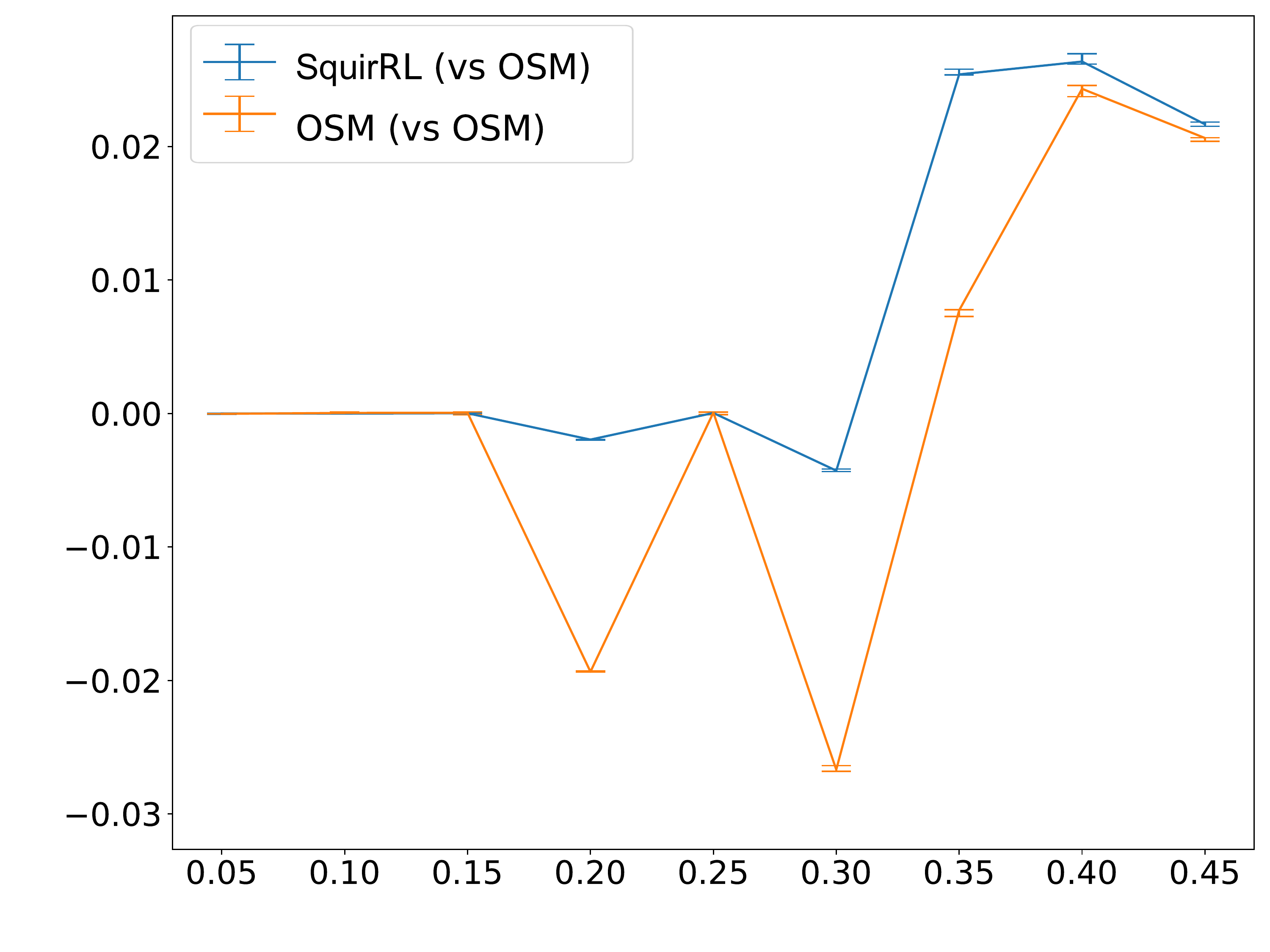}
		\put(-165,15){\rotatebox{90}{Relative reward - $\alpha$}}
		\put(-120,-10){Attacker hash power $\alpha$}
		\caption{
			\name{} gives equal or higher relative rewards compared to OSM; however, excess relative rewards can be \emph{negative}.
		}
		\label{fig:osm_vs_rl_1}
	\end{minipage}
	~~
	\begin{minipage}[b]{0.33\linewidth}
		\centering
		\includegraphics[width=0.9\textwidth]{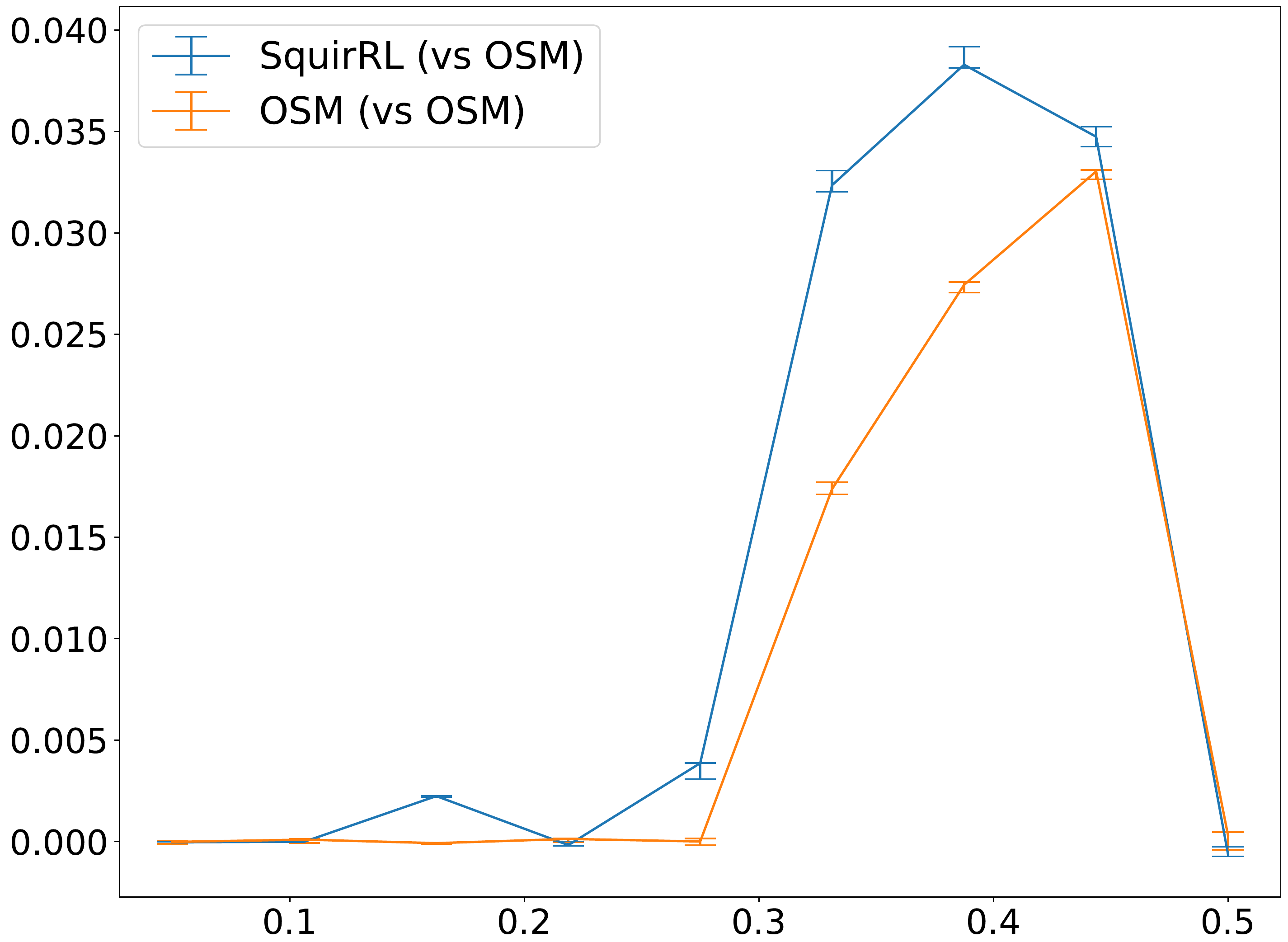}
		\put(-165,15){\rotatebox{90}{Relative reward - $\alpha$}}
		\put(-120,-10){Attacker hash power $\alpha$}
		\caption{We can alleviate the problems with the rushing adversary (\Cref{fig:osm_vs_rl_1}) by using our more realistic model. }
		\label{fig:osm_vs_rl}
	\end{minipage}
    ~~
    \begin{minipage}[b]{0.33\linewidth}
        \centering
            \includegraphics[width=0.9\textwidth]{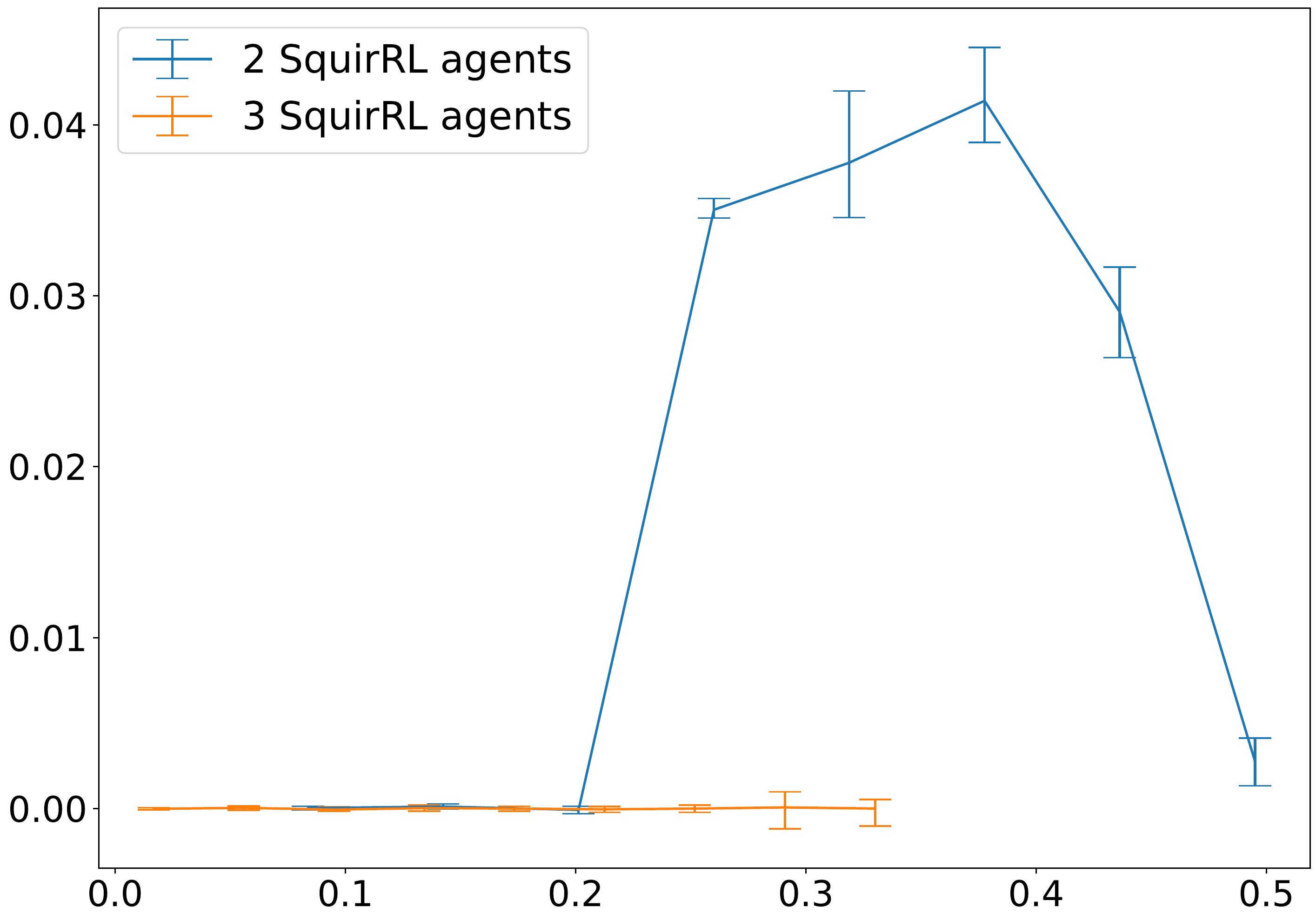}
            \put(-165,15){\rotatebox{90}{Relative reward - $\alpha$}}
            \put(-125,-10){Per-attacker hash power $\alpha$}
            \caption{
            In the multi-agent setting, we observe no gains from selfish mining in the presence of three parties.}
        \label{fig:rl_all_seg}
    \end{minipage}
\end{figure*}



\noindent \textbf{Training Methodology.}
As a starting point, our models are trained using PPO using the default configurations in \cite{rllib} for 532000 episodes, with a batch size of 524288 steps.  We will detail the training methodology inline for later experiments.  This large batch size is typical of RL applications, such as OpenAI Five \cite{dota}.
Each episode in our experiments consists of 100 block creation events.  Each error bar depicts the largest, middle, and smallest of 3 data points. Each of these data points is the average of 100,000 episodes. 


\subsection{Challenges of modeling a rushing adversary}
\label{sec:rushing}
\label{sec:osm-vs-rl}

We start by asking if OSM is a NE for multiple strategic agents. 
If OSM \emph{is} a NE,  we don't need \name{}---learning the best strategy for a single strategic agent is enough, at least for Bitcoin. 
Consider three agents:  A is honest, B is running OSM,  and  C is using DRL.
We compare this to a setting where both agents B and C are using OSM. 
Note that this setup encompasses settings where A has no hash power.

Figure \ref{fig:osm_vs_rl_1} shows Agent C's excess relative rewards (an agent's relative reward minus its hash power) when agents B and C each have a fraction $\alpha\leq 0.5$ of the hash power, and the honest agent has a $1-2\alpha$ fraction; 
we use follower fractions $\gamma_B=\gamma_C=0.5$.
Agent C always does better with \name{} than with OSM, suggesting that OSM is \emph{not} a NE with multiple agents. 
However, \Cref{fig:osm_vs_rl_1} exhibits two surprising features that require further examination before drawing such conclusions. 

First, notice the non-smoothness in Figure \ref{fig:osm_vs_rl_1}. This is an artifact of the OSM strategies solved in \cite{sapirshtein2016optimal}. 
For hash powers between 0.15 and 0.3, OSM learns one of multiple strategies that are functionally equivalent in the single-agent setting. Due to randomness in policy iteration, the solver may choose different strategies for different hash powers, but they all have the same relative reward  \cite{sapirshtein2016optimal}.
In our setting, these choices are not equivalent, and can lead to different rewards for the OSM agent. 
This effect causes the non-smoothness in \Cref{fig:osm_vs_rl_1} (also see \Cref{app:multi-agent-adaptive}), but does not indicate incorrect results.

Second, \Cref{fig:osm_vs_rl_1} shows a counterintuitive effect: the excess relative rewards of \name{} can be \emph{negative}; e.g., at hash power $0.2$, our agent performs slightly worse than the honest agent.
Although we cannot guarantee SquirRL's optimality, we observe that even if our agent uses the honest strategy, its rewards are still lower than those of the honest agent. 

This is happening because to model a \emph{worst-case} rushing adversary, we implemented the honest agent as part of the environment, as is standard in prior work \cite{eyal2018majority,sapirshtein2016optimal,semiselfish}. 
Even if the strategic agent uses the honest strategy, it is constrained to choose its actions after the honest agent, which leads to diminished rewards.
Therefore, unlike in the single-agent setting, a rushing adversary in a multi-agent setting can actually perform \emph{worse} than the honest party!





Although this phenomenon may seem like an artifact of our timing model,  we find that it applies more generally to multi-agent games with incomplete information.
Consider a game between A, B, and C,  where each player must vote for an option in set $\{0,1\}$, and agent A (our ``honest" agent) always votes randomly.
Suppose B and C know A's vote $v_A$ (i.e., a rushing model), and suppose C's final reward is equal to the number of votes for the winning option, i.e. $\max_{i \in \{0,1\}} \sum_{x \in \{A,B,C\}} \mathbbm 1 \{v_x = i\}$.
Now if B employs a strategy where it always votes for $1-v_A$,
C's reward is always 2. 
On the other hand, if A's vote is not visible to the other agents, then C's expected reward under an optimal strategy is strictly larger; with probability $1/4$, all agents will choose the same option, giving a reward of 3.

Hence, in general, a rushing adversary can lead to a strictly lower expected reward for one of the agents.
Coming back to the selfish mining setting, this suggests that the rushing adversary model we (and others) posed may not be appropriate for multi-strategic-agent settings.  An attacker should be able to mimic honest behavior perfectly, but the rushing adversary model does not allow attackers in the multi-agent setting to do that.
This observation could be of broader interest to the security community as the empirical and theoretical analysis of multi-agent systems becomes more widespread \cite{securitygames, securitygames2}.

\paragraph{Solutions}
Our findings suggest that honest agents should not necessarily be implemented as part of the environment.  Moving them outside the environment at least allows strategic parties to mimic honest strategies as they are defined within the model.
However, simply moving the honest agent out of the environment poses new challenges, by preventing agents from being able to react to honest actions. In reality, block propagation times are generally much faster than block mining speeds \cite{propagation}, so strategic miners should have time to react to published blocks before the next block is released.  Not allowing this makes it difficult to extract excess rewards.  



To incorporate both constraints, we make two modeling choices that are a notable departure from prior literature (1) we model the honest party as an agent with a fixed strategy outside the environment (2) instead of assuming a block is mined at every time slot, we have a block mining event every $m$ turns (in our experiments, we let $m = 4$).  
The honest party will always act in the turn following a block mining event, giving attackers time to react before the next block is mined.  We refer to this as the "time-segmentation model".
This changes the POMG: the state space $S$ now includes "time", and so does the observation space $\Omega$.  Furthermore, $O^i(s)$, the observation agent $i$ sees at state $s$ also includes the time.  

\noindent \textbf{Training Methodology.}
The more realistic model introduces a substantial new difficulty to the training process: a longer time horizon over which to optimize.  With sparser block mining events, the agent must learn to \emph{plan}.  This is a widely-acknowledged difficulty in DRL \cite{dota}, and simply running the default PPO configuration from RLlib produces poor results.  To combat this difficulty, we leverage the existing structure in the problem to modify the training methodology. 
Our modifications are as follows:

\begin{itemize}[align=left]
	\item Train for longer: approximately 2M episodes
	\item Anneal $m$ from 0 to the desired value of 4, increasing $m$ by 1 every 500K steps.
	\item We want to detect vulnerabilities, so we bias the agents towards selfish mining by adding a bonus of $0.1 \cdot \textrm{max}(2\textrm{M} - \textrm{total episodes}, 0)/2\textrm{M}$ if the agent waits between episode 500K and 1.6M.  In general, biasing agents towards dishonest behavior is a good choice when analyzing the security of a system.  
	\item At environment initialization, we run OSM agents in place of \name~agents in the game for a random number of block creation events.  Then we use this leftover state as the initial episode state.  It is not necessary to use the OSM strategy for initialization; any initialization of states that gives sufficient coverage over all states suffices \cite{distributionshift}.
	\item Set the discount factor $\eta = 0.997$ rather than $\eta = 0.99$ to increase the incentive for the agent to plan ahead.
	\item Batch size of 1048576 steps.
\end{itemize}  

In Sections \ref{sec:osm-ne} and \ref{sec:selfish-k3}, we demonstrate that we obtain physically realistic results under our new modeling choices, in addition to obtaining other novel results.

 

\subsection{OSM  is not a Nash equilibrium}
\label{sec:osm-ne}


We apply this new time-segmented, non-rushing model to obtain Figure \ref{fig:osm_vs_rl}. 
We now observe the more physically realistic result that when Agent C is instantiated with SquirRL or OSM, it always outperforms honest agents, unlike in the previous model. Notice that as $\alpha \to 0.5$, the excess relative reward tends to zero because the honest party's hash power tends to 0, so there is less excess reward to claim.

Furthermore, in a competition with OSM, agent C does better using DRL (blue line) than OSM (orange line). This implies that OSM is not actually a NE.  In other words, the approach of \cite{semiselfish} to analyze restricted strategy sets is not sufficient.  If we had restricted agents to either honest mine or to follow OSM, then we might have (incorrectly) concluded from Figure \ref{fig:osm_vs_rl}'s orange line that OSM is a NE, similarly to how \cite{semiselfish} concludes that semi-selfish mining is a NE.

\subsection{Selfish mining may be unprofitable with $k\geq 3$ agents}
\label{sec:selfish-k3}

\begin{figure}[t]
    \centering
		\includegraphics[width=0.85\columnwidth]{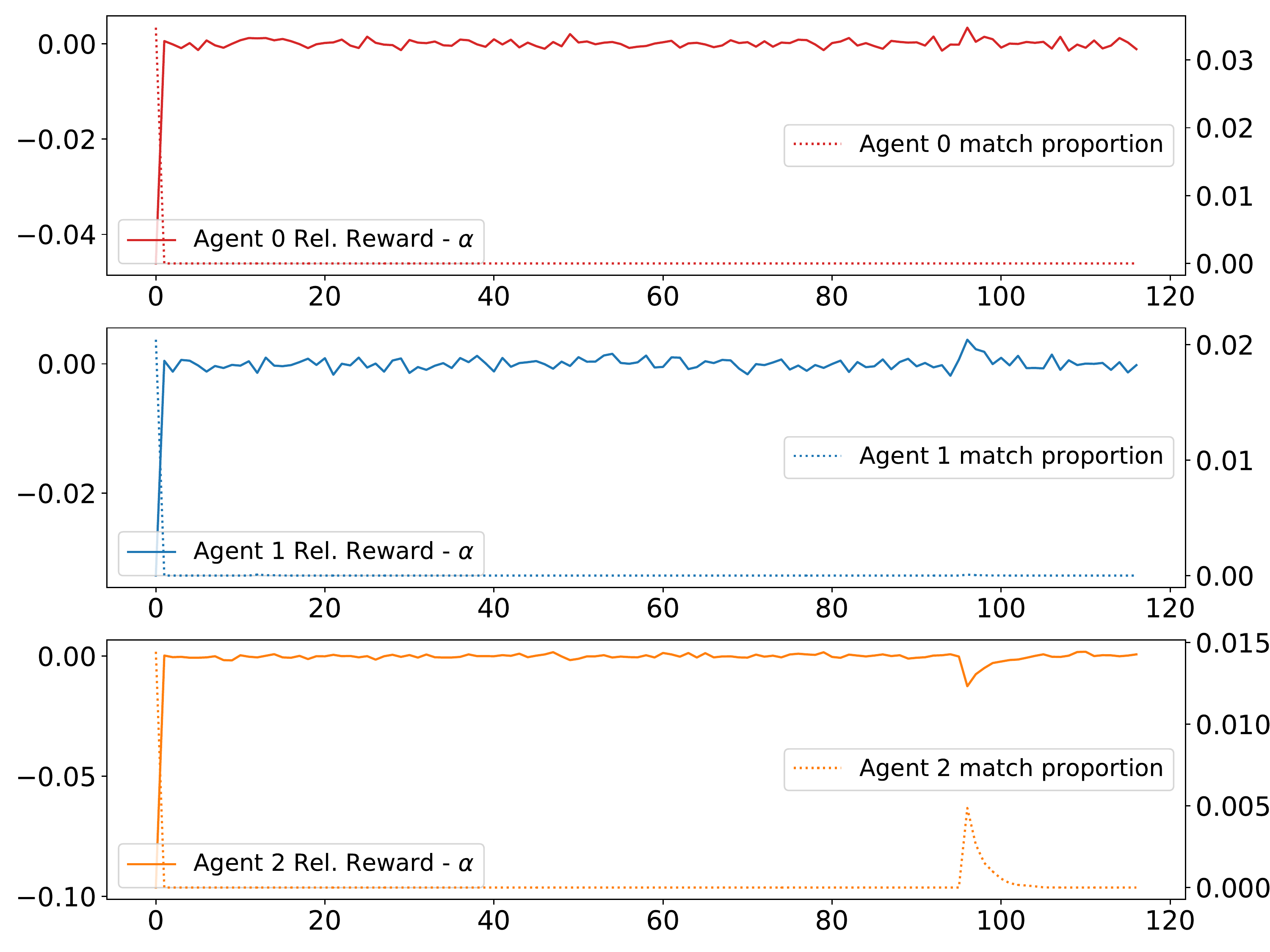}
		\put(-145,-10){Iteration Number}
		\put(-225,40){\rotatebox{90}{Relative Reward - $\alpha$}}
		\put(5,28){\rotatebox{90}{Fraction of `match' actions}}
        \caption{Deviations from the equilibrium strategy are short-lived:~\name~Agent 2 chooses to increase its matching proportion at around iteration 100, loses reward immediately, and then returns to the equilibrium strategy.  These curves are at 0.1733 hash power for each strategic agent, with the rest given to the honest agent.  $\gamma_i = 1/3$ for all strategic agents.}
    \label{fig:rl3_iterations}
\end{figure}

Our next experiments involve training multiple strategic agents against one another in a selfish mining game under the time-segmented, non-rushing model. 
Notice that these experiments can \emph{only} be run with DRL, as the environment is both unknown and dynamic. We highlight one observation from Figure \ref{fig:rl_all_seg}: with three adaptive strategic agents, the agents could not achieve reward better than honest mining.  

Notice that the training modifications we detailed in Subsection \ref{ssec:model} all bias our agents to behave more selfishly.  In addition, we let $\gamma_i = 1/3$ for all $i$, the maximum possible follower fraction: however, the equilibrium the agents settle on is honest mining.  Figure \ref{fig:rl3_iterations} illustrates in solid lines the relative reward of each agent minus its hash power ($\alpha = 0.1733$) and in dotted lines the fraction of match actions, which is a proxy for the agent's strategy. Matching more often is a more aggressive strategy; the honest strategy never matches. When agents deviate from the honest strategy by matching more (e.g., agent 2 around iteration 100), they lose reward and quickly revert to an honest strategy.  These experiments suggest (but do not prove) that honest mining is a Nash equilibrium for $k\geq 3$ strategic agents.


\begin{figure*}[t]
	\begin{minipage}[b]{0.32\linewidth}
		\centering
		\includegraphics[width=\textwidth]{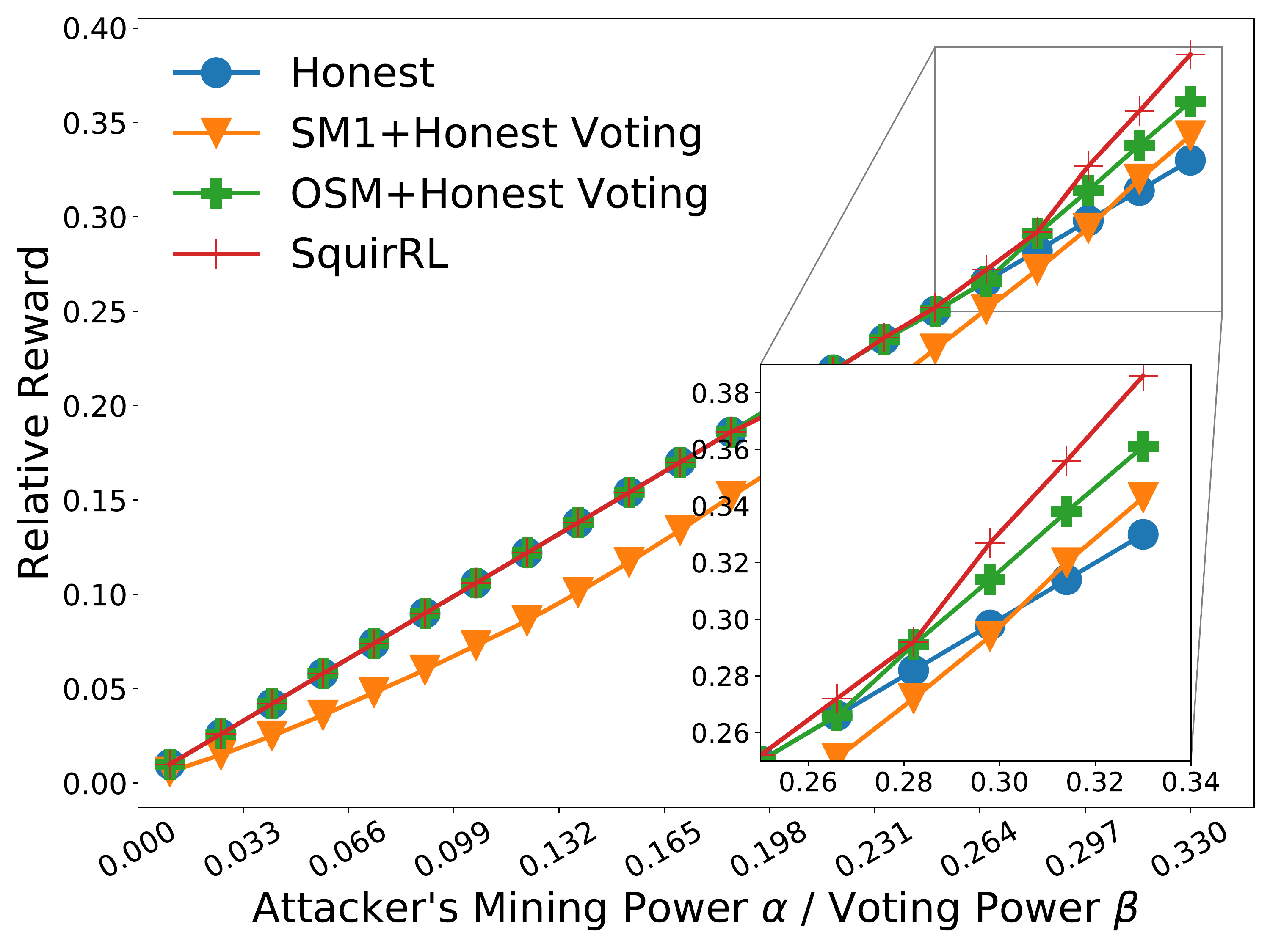}
		\caption{Total relative reward, including both voting and mining rewards.}
		\label{fig:casper_total}
	\end{minipage}
	~~
	\begin{minipage}[b]{0.32\linewidth}
		\centering
		\includegraphics[width=\textwidth]{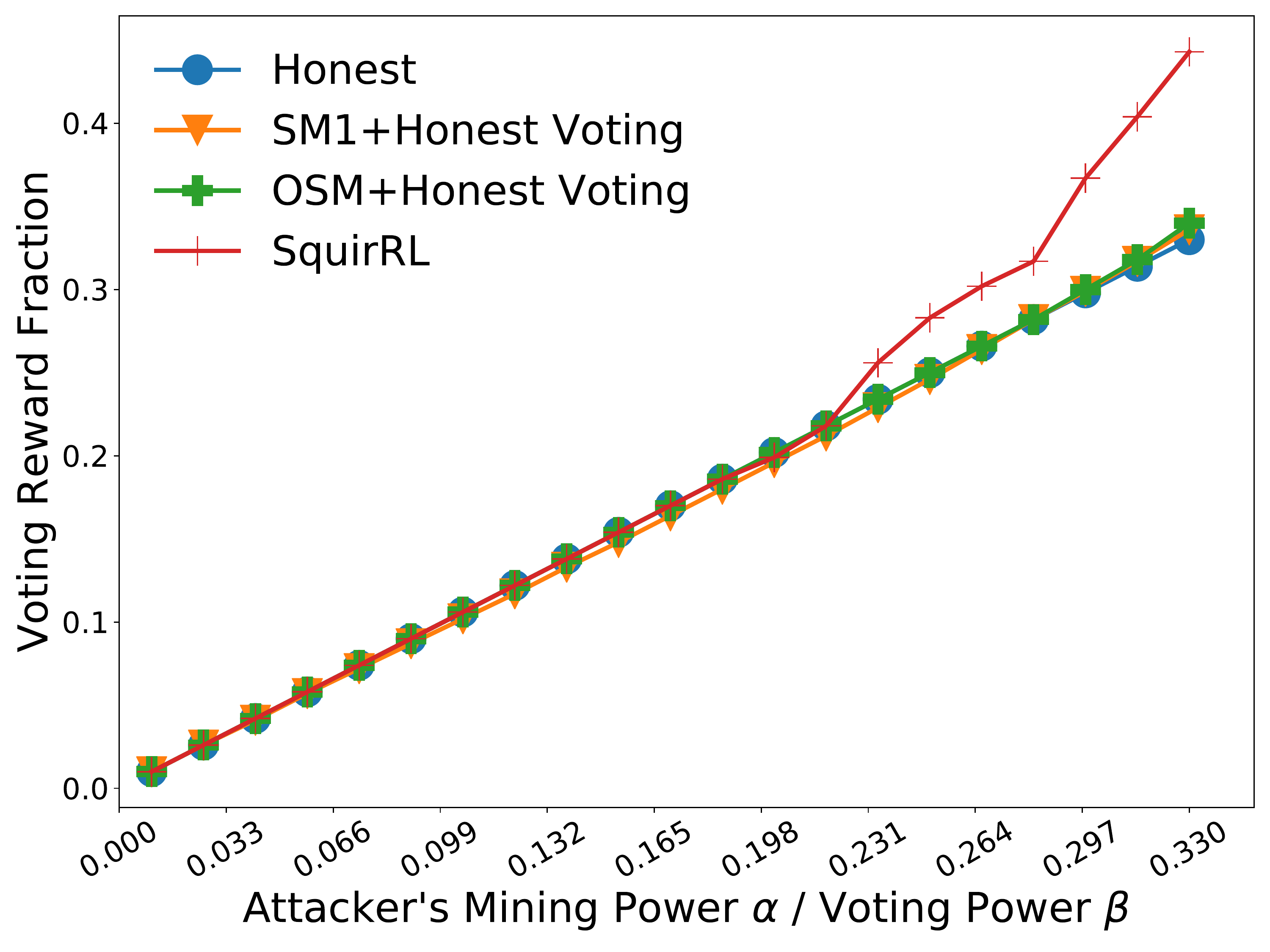}
		\caption{Relative reward from Casper FFG (not including mining rewards).}
		\label{fig:casper_voting}
	\end{minipage}
	~~
    \begin{minipage}[b]{0.32\linewidth}
        \centering
        \includegraphics[width=\textwidth]{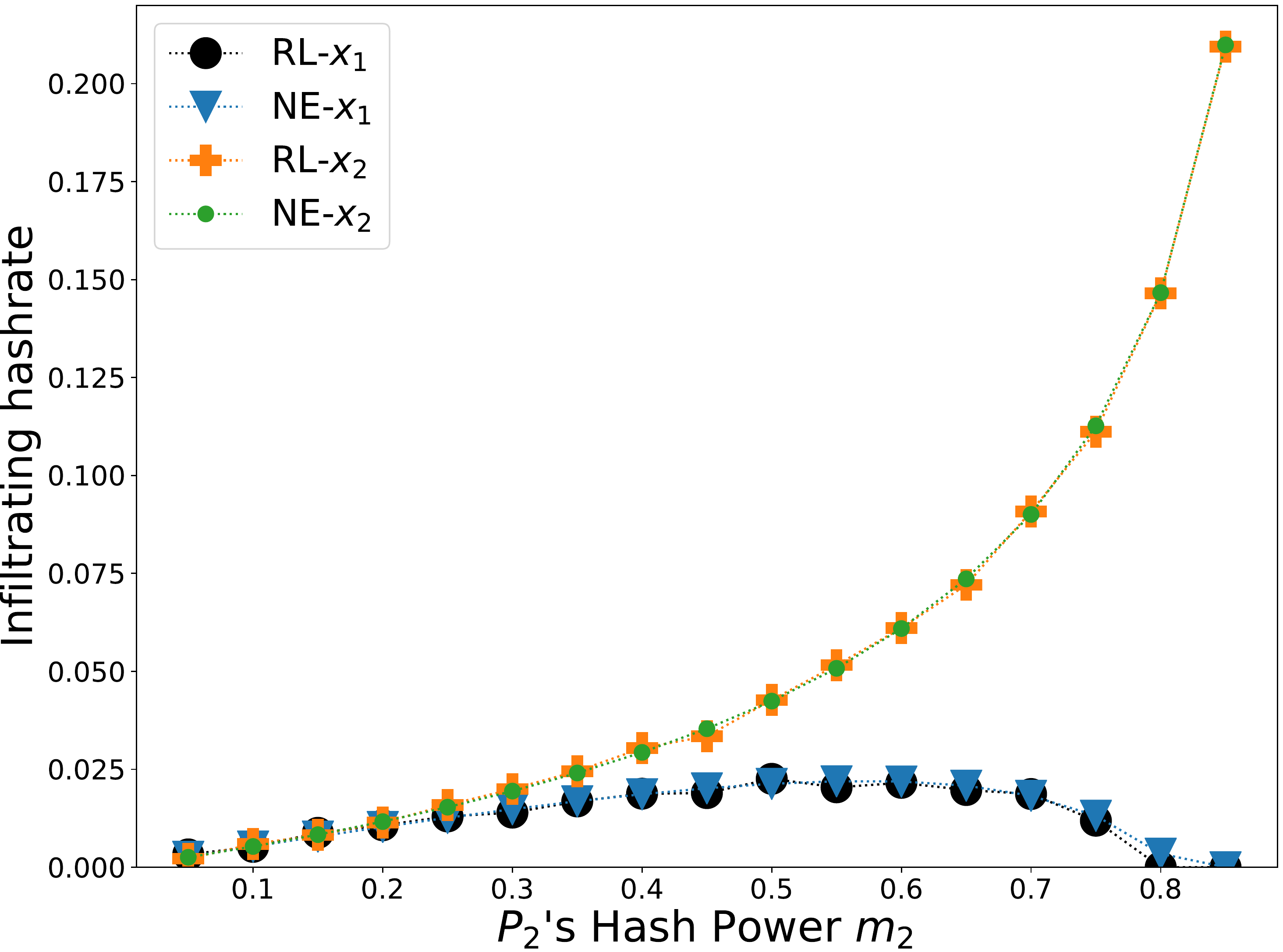}
        \caption{Policies of NE vs RL. $P_1$'s hash rate $m_1 = 0.1$, and $P_i$ uses hash power $x_i$ to infiltrate the other pool.
        }
            \label{fig:blockwithholding_policy}
    \end{minipage}
\end{figure*}

\section{Beyond Selfish Mining}
\label{sec:case-studies}
We have thus far focused on selfish mining attacks. 
To show the general applicability of \name{}, we apply it to two problems that are not selfish mining: voting-based finality protocols \cite{casper} and block withholding \cite{eyal2015miner}.

\subsection{Casper the Friendly Finality Gadget (FFG)}
\label{sec:casper}
In this section, we demonstrate a novel attack on the Ethereum blockchain's planned finalization protocol, called 
Casper the Friendly Finality Gadget (FFG) \cite{casper}. 
Casper FFG is a proof-of-stake (PoS), voting-based protocol for finalizing blocks in proof-of-work (PoW) blockchains. 
Casper FFG includes an incentive mechanism to ensure that participating nodes, or \emph{validators} do not deviate from the desired behavior. 
Our goal is to use \name{} to exploit these incentive mechanisms to amplify an agent's reward and/or subvert the integrity of the voting process. 
Our attack illustrates how an agent can \emph{combine} PoW selfish mining with PoS strategic voting to amplify her own rewards.
To the best of our knowledge, this is the first attack to combine selfish mining with strategic voting in Byzantine fault-tolerant (BFT)-style protocols. 
These experiments \emph{cannot} be solved with value/policy  iteration because the state space is continuous.


Casper FFG validators  are meant to finalize the first block
(or \emph{checkpoint}) of every \emph{epoch}, defined as a chunk of consecutive $\ell$ blocks on the same chain. Finalization occurs via voting. When a checkpoint receives more than 2/3 of the votes, it is \textit{justified}. Here, the votes are weighted by the voters' deposits. If multiple checkpoints exist at the same height, a validator should vote for the checkpoint on the longer chain. 
If two consecutive checkpoints on the same chain are justified, the first checkpoint is \textit{finalized} and it will remain in the canonical chain forever. 
A chain that does not include every finalized checkpoint in the system is considered invalid, even if it is the longest (greatest-work) chain in the system.

The Casper FFG incentive mechanism is designed to ensure that validators (a) participate in every epoch's voting protocol, and (b) vote for the same checkpoint if  multiple options exist. 
To achieve this, each validator $v$ makes a deposit $D_v$ into a smart contract on the Ethereum chain to join the validator pool. 
Roughly, if a checkpoint $c$ at height $h(c)$ is finalized, then all the voters who voted \emph{for} $c$ see their deposit grow, whereas any voters who voted for a different checkpoint $c'$ with $h(c')=h(c)$ will see their deposit shrink.

We implement a simplified version of the Casper FFG incentive mechanism and voting process that captures the essence of the protocol.  
At any given time step, we model the voting process as either \textit{active} or \textit{inactive}.
If the voting process is \textit{inactive}, then the actions and state transitions are the same as in selfish mining.  
If the voting process is \textit{active}, then the attacker can choose: (1) to vote, in which case it allocates all of its votes to its own fork, (2) to publish blocks, in which case the transitions mimic those in the selfish mining setting, or (3) to wait and continue mining without taking any publicly-facing actions.  
While the attacker is waiting, honest validators can vote for a checkpoint. 
In practice, validators will not vote simultaneously; we model this heterogeneity by staggering the honest votes according to a random process. 
In a given time slot, with probability $p_{\textrm{vote}}$ the honest parties randomly choose a longest chain to vote for.  We allocate to this longest chain a proportion of honest votes equal to  $\min(\max(X, 0), 1 - \beta - v_h)$, where $X \sim \mathcal N(0.1, 0.05)$, $v_h$ is the total proportion of votes already allocated by the honest parties, and $\beta$ is the proportion of total votes available that are under the control of the attacker.  
In other words, we choose a random fraction of the remaining, uncast honest votes and allocate them to the checkpoint on the current longest chain.
The random vote allocation distribution reflects the non-uniformity of block propagation in the blockchain.  
With probability $1 - p_{\textrm{vote}}$, a block is generated by a miner and the block structure changes according to the selfish mining setting.
The voting process is active at the beginning of any epoch.  It becomes inactive if (1) one checkpoint receives more than $2/3$ of the votes (2) one chain containing a checkpoint becomes the canonical chain through selfish mining transitions---for instance, if the attacker chooses to adopt the honest chain.

Block rewards are calculated as follows: miners get one unit of reward for every block that ends up on the canonical chain.  Voting rewards are calculated (roughly) as in Casper FFG, with the validator deposit $D_v$ scaled appropriately to reflect the ratio of real-world deposit magnitudes to block rewards. The differences between our modeling assumptions and the full scheme are detailed in \Cref{sec:casper_experiment}.

\noindent \textbf{Results.}
As the agent's voting and mining power increases, \name{} learns to exploit the penalty for incorrect voting to penalize the honest party. A common attack strategy discovered by \name{} is as follows.  First, the~\name~agent accumulates and hides two checkpoints, $c' \to c''$, through selfish mining. When the honest party releases a checkpoint $c$, the agent immediately releases $c'$ and triggers a competing voting process. The agent then waits until the honest checkpoint $c$ accumulates close to (but not above) 1/3 votes. The agent then releases $c''$, causing checkpoint $c'$ to be included in a longer chain than $c$. The remaining voters will vote for $c'$ according to the voting rule and $c'$ will be justified. The honest voters for $c$ are penalized, amplifying the agent's relative reward.

Figure \ref{fig:casper_total} shows the total relative rewards accumulated by an agent with the same fraction in mining hash power and voting pool deposit. We vary this fraction up to 1/3 because Casper FFG is not secure above 1/3 adversarial voting power, but honest voting is shown to be a Nash equilibrium for agents with $<1/3$ voting power in \cite{buterin2019incentives}.\footnote{This result considers Casper FFG voters in isolation, without accounting for the possibility of an agent's concurrent mining activity.}
We observe that this attack allows the agent to amplify its rewards more than selfish mining alone. Figure \ref{fig:casper_voting} illustrates dramatic gains in relative voting reward, up to 30\% over honest rewards. 

This attack raises an important practical concern. 
The interest rate associated with voting rewards in Casper is close to extrinsic interest rates (e.g. the stock market). 
Hence, if an adversary is able to drive down an honest participant's rewards, honest voters may leave the voting pool, making it easier for an adversary to control more than 1/3 of the voting power. 
Hence, this attack can actually affect the integrity of the finalization mechanism itself. 
An important implication of this case study is that system designers should consider how incentive mechanisms \emph{compose} with other incentive mechanisms. That is, Casper FFG in isolation is not vulnerable to these attacks. It is only by combining mining incentives with the Casper FFG voting protocol that we observe this vulnerability. 

\subsection{Block Withholding Attacks}
\label{ssec:block-withhold}


We explore a second case study in which agents perform block withholding attacks~\protect{\cite{rosenfeld2011analysis, eyal2015miner, laszka2015bitcoin, luu2015power, haghighat2019block}}, an attack observed in practice (see, e.g.,~\protect{\cite{BlockWithholdingCase}}). 
In block withholding attacks, a mining pool infiltrates miners into opponent pools to diminish their revenue and gain a competitive advantage. 
The attacking pool deploys mining resources in a target pool and submits partial solutions, i.e., proofs of work, to earn rewards. 
If the attacking pool mines a block in the target pool, it withholds it. The target pool thus loses block rewards and revenue relative to its hash power declines.

In prior work, Eyal~\cite{eyal2015miner} showed that for two competing mining pools, there is a (unique) Nash equilibrium where each pool assigns a fraction of its resources to infiltrate and sabotage the other. \name~automatically learns pool strategies that converge to the same revenues as predicted by that equilibrium.

We adopt the same model as in Eyal~\protect{\cite{eyal2015miner}}. 
In the two-party version of  this model, strategic mining pools $P_1$ and $P_2$ each possess ``loyal'' miners with hash rates of $m_1$ and $m_2$, respectively, $0 \leq m_1 + m_2 \leq 1$. 
The remaining miners mine on their own, not forming or joining a pool. 
A miner loyal to pool $P_i$ may either mine honestly in $P_i$ or infiltrate $P_{3-i}$, as dictated by $P_i$. 
When an infiltrating miner loyal to $P_i$ generates a  partial  block reward, the reward is relayed to $P_i$ and split among all registered miners in $P_i$, as well as the miners who are loyal to $P_i$ but currently infiltrating $P_{3-i}$. 
The goal is to maximize the revenue of each miner, normalized by the revenue when there is no block withholding attack. 

Denote the hash power of miners loyal to $P_i$ and infiltrating $P_{3-i}$ by $x_i$. Thus $0 \leq x_i \leq m_i$. We set up the two-agent RL experiment using the reward functions defined in~\protect{\cite{eyal2015miner}}. 
Each agent is assigned a mining hash power $m_i$ and aims to maximize its reward by choosing $x_i$, the hash power infiltrated into the other pool, from a continuous action space $[0, m_i]$.
The reward to be optimized is the immediate normalized revenue (again, as defined in~\protect{\cite{eyal2015miner}}).
There is no state transition in this environment: the game has episode length 1.
Two agents take turns to adapt their strategies given the best strategy the other agent learned in the last episode.
We trained the model using PPO because it is suitable for the multi-agent setting, as mentioned before, and supports continuous action spaces.

For reproducibility, our setting of hyperparameters in PPO and training results are specified as follows. 
We set the clipping parameter $\epsilon$ 
to $0.1$, with a linear learning rate schedule decaying from $10^{-5}$ to $10^{-7}$.
The entropy coefficient $\beta$ is set to $0.01$ initially and decays to $\beta \leftarrow \beta  (1 - timestep/schedule)$ every training step, with $schedule$ set to $10^{9}$.
After $10^6$ episodes, both the strategies and rewards converge to those in the Nash equilibrium specified in~\protect{\cite{eyal2015miner}}, to within 0.01.
The detailed policies and revenues are plotted in~\cref{fig:blockwithholding_policy,fig:blockwithholding_reward} respectively. 

Related work in~\cite{haghighat2019block} uses
RL, specifically a policy gradient based learning method~\protect{\cite{bowling2002scalable}}, to study block withholding among multiple agents in a setting with dynamic hash rates. A limitation of that work is that it uses a discrete action space, not a continuous one. One interesting feature is its inclusion of a probabilistic model of migration of unaffiliated (free agent) miners to the most successful pools, an extension of the model in~\cite{laszka2015bitcoin}. Unfortunately, this model is rather artificial, with no grounding in empirical study, so we chose not to duplicate it.

\section{Related Work}
\label{sec:related}

A number of recent works have analyzed direct attacks on and economic flaws in cryptocurrency protocols.

\noindent \textit{(1) Selfish Mining.}
The concept of selfish mining was introduced by Eyal and Sirer in~\protect{\cite{eyal2018majority}}, and a large body of resulting work has sought to refine related mining models and compute protocol security thresholds in a selfish mining context\protect{~\cite{sapirshtein2016optimal, nayak2016stubborn, niu2019selfish, ritz2018impact}}.   Much of this work (including\protect{~\cite{sapirshtein2016optimal}}) uses MDP solving to compute optimal selfish mining strategies.  These exact solutions are less robust to unexpected, real-time changes in honest hashpower than our RL-based approach.  
An enhanced model with two selfish agents and one honest agent is considered in\protect{~\cite{bai2018deep}}, but this work does not consider the presence of multiple rational actors in the network, which is more realistic for a cryptocurrency mining setting.  
Our techniques allow for reasoning about more realistic models at the expense of theoretical guarantees.  

\noindent \textit{(2) General Mining Attacks.}
A wide range of work has also focused on potential mining attacks besides selfish mining. 
One example is difficulty attacks~\Cite{meshkov2017short,goren2019mining}, in which miners can profitably manipulate a chain's difficulty by secretly raising difficulty on their own private chain~\Cite{bahack2013theoretical}, switching between competing currencies secured by the same mining hardware~\Cite{meshkov2017short}, or pausing mining activities around difficulty adjustment time~\Cite{goren2019mining}.  In some situations, this can discourage miners from mining at all~\Cite{blockchaindeathspiral}.  The study of inducing discouragement in our framework is left to future work.

Attacks involving miner manipulation of user transactions, via censorship or reordering, are surveyed in~\Cite{judmayer2019pay}.  Many such attacks have been shown in theory, allowing miners to double-spend user funds or profit from incorrect assumptions in second-layer applications~\Cite{bonneau2016buy,liao2017incentivizing,mccorry2018smart}.  Such attacks have been observed in practice~\Cite{eskandari2019sok,daian2019flash} and can be performed without hashpower by bribing existing miners~\Cite{mccorry2018smart,teutsch2016cryptocurrencies}.  Because these attacks yield direct revenue for miners, they can almost certainly be used to subsidize a successful mining attack as described in our work, lowering the required hashpower threshold.

Lastly, attacks against miners are possible at the network layer.  One example is DoS attacks on mining pools, which have been observed in practice~\Cite{vasek2014empirical} and which more often affect larger pools~\Cite{johnson2014game}.  Another is eclipse attacks~\Cite{castro2002secure,sit2002security,singh2006eclipse}, which can ensure a node is connected to only attackers and is applied to blockchain systems in~\Cite{heilman2015eclipse,marcus2018low}.  It has shown that such attacks can interact with selfish mining to increase efficacy~\Cite{nayak2016stubborn}, and routing-based eclipse attacks have been observed in blockchains~\Cite{BGPhijacking}.

\noindent \textit{(3) RL, MDPs, and Computer Security.}
Our work focuses in part on analyzing multi-agent games with DRL.  A connection between MDPs and mult-agent RL was made in a seminal paper by Littman~\Cite{littman1994markov}, which proposes the use of RL to extend MDP analysis to multi-agent games.  Recent work (e.g.,~\Cite{shiva2010game,wang2019deep,yu2018deep}) has applied this technique to cybersecurity actors, analyzing meta-games between attackers and defenders.  Our work extends this style of analysis into a setting where multiple ``attackers" compete to maximize their own profit share.  Unlike in traditional security, our setting is not mutually exclusive (multiple attackers can profit), and requires attackers to continually respond to each others' actions.  This greatly increases strategy space complexity in a way likely inherent to the open participation of most cryptocurrency protocols.
\section{Conclusion}
\label{sec:conclusion}

In this work, we propose \name, a deep RL-based framework to automate vulnerability detection in blockchain incentive mechanisms. 
We have shown that \name~can approximate known theoretical results regarding attacks on blockchain incentive mechanisms. It can also handle settings that are intractable using classical techniques like policy iteration, such as multiple competing agents or continuous state spaces.

\name{} cannot prove the security of a mechanism.   
We have shown, however, how \name~can serve as a powerful ``quick-and-dirty" tool, allowing protocol designers to gain intuition about their protocols in cases where theoretical  analysis is infeasible. We also believe that future work will illuminate new uses,
including other classes of incentive-based attacks, e.g., time-bandit attacks \cite{daian2019flash}.

\bibliographystyle{plain}
\bibliography{ref} 

\appendix

\subsection{Relative vs. Exact Rewards}
\label{app:near}

The \textbf{difficulty} of a block is the average computation required to mine it. 
If the computational power of the whole network increases or decreases, the block difficulty increases or decreases accordingly to maintain a target block generation time  under dynamic network conditions. 
For instance, in Bitcoin, the target main chain block generation rate, a.k.a. growth rate, is $T_0=10$ minutes/block, and the difficulty is adjusted every $M=2016$ blocks; we call this duration an \emph{epoch}. 
The difficulty $d'$ of one next epoch is adjusted according to the current difficulty $d$ and the average main chain growth rate $T$ of the current epoch, such that $d'=d\frac{T_0}{T}$. 

We next show that over $n$ epochs with difficulty adjustment, the absolute reward rate (scaled by a constant) and relative reward are close  for a a strategic agent against an arbitrary number of other agents.
In what follows,  we let $S_a$ and $S_o$ denote the number of stale blocks mined by the adversary of interest and all the other miners in a single epoch, respectively; a stale block is a block that does \emph{not} end up on the main chain, and hence does not collect any block reward. 
Note that the following result relies on the deterministic analysis of \cite{goren2019mining},  which abstracts away the stochastics of the problem. 
\begin{figure}[t]
    \centering
    \includegraphics[width=0.8\columnwidth]{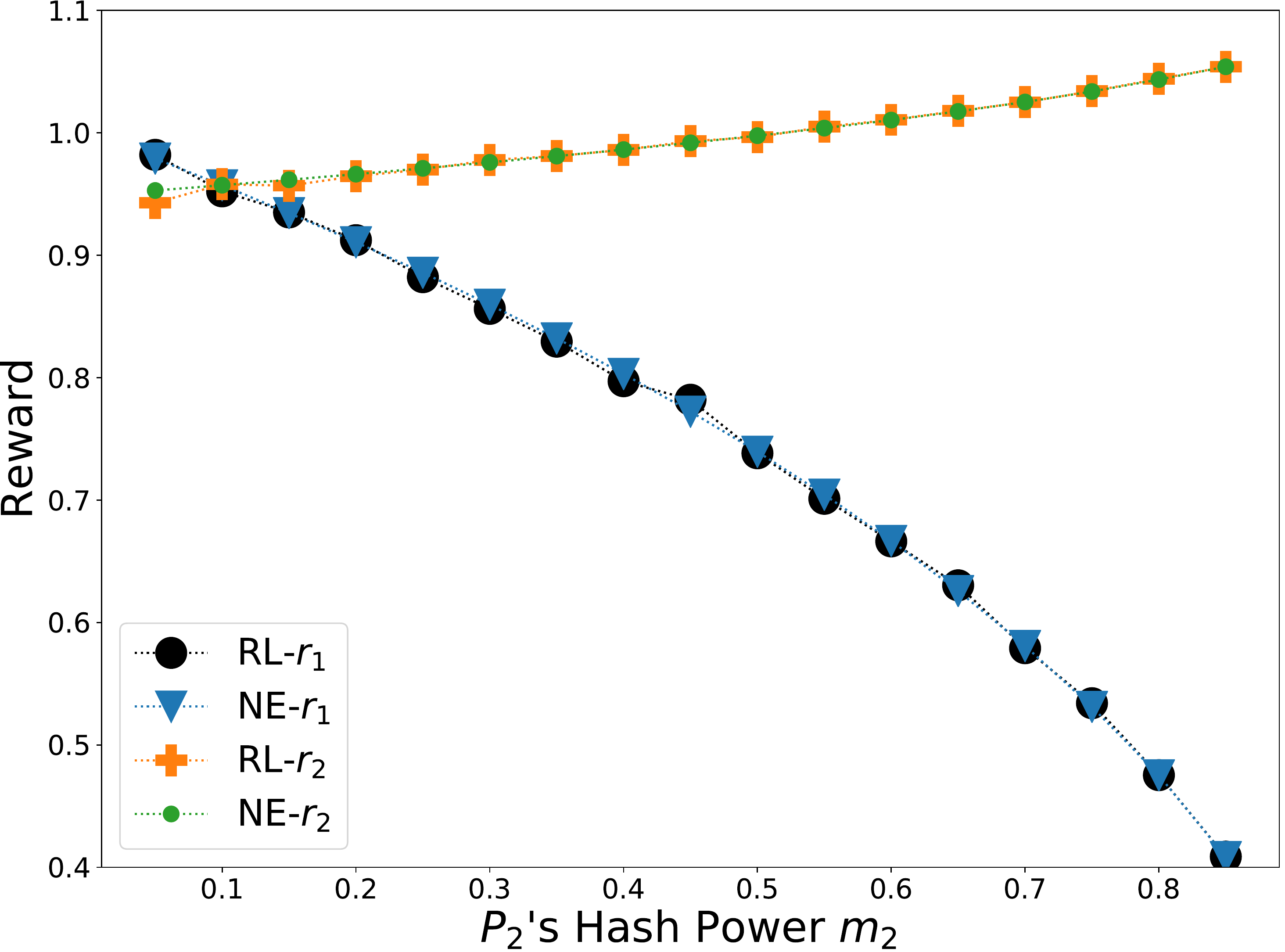}
    \caption{Rewards of NE vs RL. $P_1$'s hash rate $m_1 = 0.1$. Here, $r_1$ denotes the reward of $P_1$ and $r_2$ that of $P_2$.}
        \label{fig:blockwithholding_reward}
\end{figure}
\begin{prop}
Let $R_n$ denote the expected absolute reward rate over $n$ epochs,  $\tilde R_n$ the expected relative reward rate, $k$ the number of independent actors (all honest parties can be considered to be one actor), and $T_0$ the target (expected) inter-block time. 
Consider an attacker with $\alpha < 0.5$ fraction of the total hash power in the network.
We assume that 1) the attacker always uses all its mining power, 2) the attacker uses the same strategy across all epochs, and 3) the total hash power of the network remains unchanged across epochs. 
Then under a deterministic analysis model, 
\begin{align}
R_n = \frac{1}{T_0} \frac{B_a}{(B_a+B_o)+\frac{1}{n}(S_a+S_o)},    
\label{eq:rn}
\end{align}
which in turn implies that
$
|T_0 R_n - \tilde R_n| \leq \frac{k-1}{n}.
$
\label{prop:near}
\end{prop}
The proof can be found in \cref{app:near_proof}.
Proposition \ref{prop:near} has two main implications. 
The first is that over the course of a single epoch, honest mining is an optimal strategy  for maximizing the absolute reward rate; selfish mining is actually less profitable.
This follows because to maximize the absolute reward in \eqref{eq:rn},  the denominator should be minimized. 
This can be achieved by producing no stale blocks, which occurs under honest mining.

The second implication is as follows: since maximizing absolute reward rate is equivalent to maximizing absolute reward rate scaled by a positive constant ($T_0$), Proposition \ref{prop:near} suggests that for moderate $n$, the objective functions for optimizing relative rewards and absolute reward rate are provably close.
In particular, we have that 
$$
\lim_{n\to \infty} R_n = \frac{\tilde R_n}{T_0},
$$
i.e., in the infinite-time horizon, optimizing absolute reward rate is equivalent to optimizing relative rewards. 
\name~can  be used to optimize both absolute rewards and relative rewards; however, to compare with prior literature and because  of this asymptotic equivalence, we  start by considering relative rewards.  

\subsubsection{Proof of Proposition \ref{prop:near}}
\label{app:near_proof}
As mentioned before, we use the deterministic analysis of \Cite{goren2019mining}. 
This analysis deals entirely with expectations, and abstracts away the randomness in block time generation; it is a good approximation when the epoch duration is high (as it is in Bitcoin). Let the number of independent actors in the blockchain be $k$.
We denote the number of main chain blocks generated by the attacker during the $i$th epoch as $B_a(i)$ (assuming $1\leq i\leq n$) and the number of stale blocks generated by the attacker as $S_a(i)$; we suppress the notation $p$ for simplicity. 
Here a stale block refers to any block that does not end up on the main chain.
Those numbers for all other parties combined are are $B_o(i)$ and $S_o(i)$. 
We have $B_a(n)+B_o(n)=M, n \in \mathbb N_+$ since every epoch has $M$ of blocks in the main chain. 

In the first epoch, the average block generation time is $T_0$, but the main chain growth rate may be lower than $1/T_0$ if the attacker deviates from the honest mining protocol. 
Therefore, the total duration of the first epoch is $D_1 = (M+S_a(1)+S_o(1))T_0$. 
After the first difficulty adjustment, the difficulty will be multiplied by $M/(M+S_a(1)+S_o(1))$, so the expected duration of the second epoch is 
$$
D_2 = (M+S_a(2)+S_o(2))T_0 \frac{M}{(M+S_a(1)+S_o(1))}.
$$ 
Recall that we assume all parties repeat their strategies for all epochs, so we have $B_a(1)=B_a(2),B_o(1)=B_o(2),S_a(1)=S_a(2),S_o(1)=S_o(2)$ under deterministic analysis \protect{\cite{goren2019mining}}. 
We therefore use the simplified notation $B_a, B_o, S_a, S_o$ and suppress notation  $n$. 
Therefore, the total time for the second epoch is actually $MT_0$. 
This pattern holds for larger $n$ by induction.
We can therefore write the absolute reward rate of the attacker for $n$ epochs $R_n$ as follows:

\begin{align}
    R_n =
    & \frac{nB_a}{(M+S_a+S_o)T_0 + (n-1)MT_0} \nonumber \\
    &= \frac{nB_a}{nMT_0+(S_a+S_o)T_0} \nonumber \\
    &= \frac{1}{T_0} \frac{B_a}{(B_a+B_o)+\frac{1}{n}(S_a+S_o)}.  \label{eq:abs}
\end{align}
Notice that optimizing the absolute reward rate for $n$ epochs $R_n$ is equivalent to optimizing $T_0 R_n$, since this just scales the objective by a constant.
Computing the difference between $\tilde R_n$ and $T_0 R_n$,  we get 
\begin{align}
    |T_0 R_n - \tilde R_n| 
    &= \frac{B_a}{B_a+B_o} - \frac{B_a}{(B_a+B_o)+\frac{1}{n}(S_a+S_o)} \nonumber \\ 
    &= \frac{B_a(B_a+B_o + \frac{1}{n}(S_a+S_o))-B_a(B_a+B_o)}{(B_a+B_o+\frac{1}{n}(S_a+S_o))(B_a+B_o)} \nonumber \\
    &\leq  \frac{M(\frac{(k-1)M}{n})}{M^2} = \frac{k-1}{n} \label{eq:diff}
\end{align}
where \eqref{eq:diff} follows because $B_a+B_o=M$ and $S_a+S_o \leq (k-1)M$, because there can only be at most $k$ branches at a time and only the longest chain ends up as the main chain.
This gives the  claim.

\begin{figure*}[!htb]
	\centering
	\begin{minipage}[b]{\columnwidth}
		\centering
		\includegraphics[width=\textwidth]{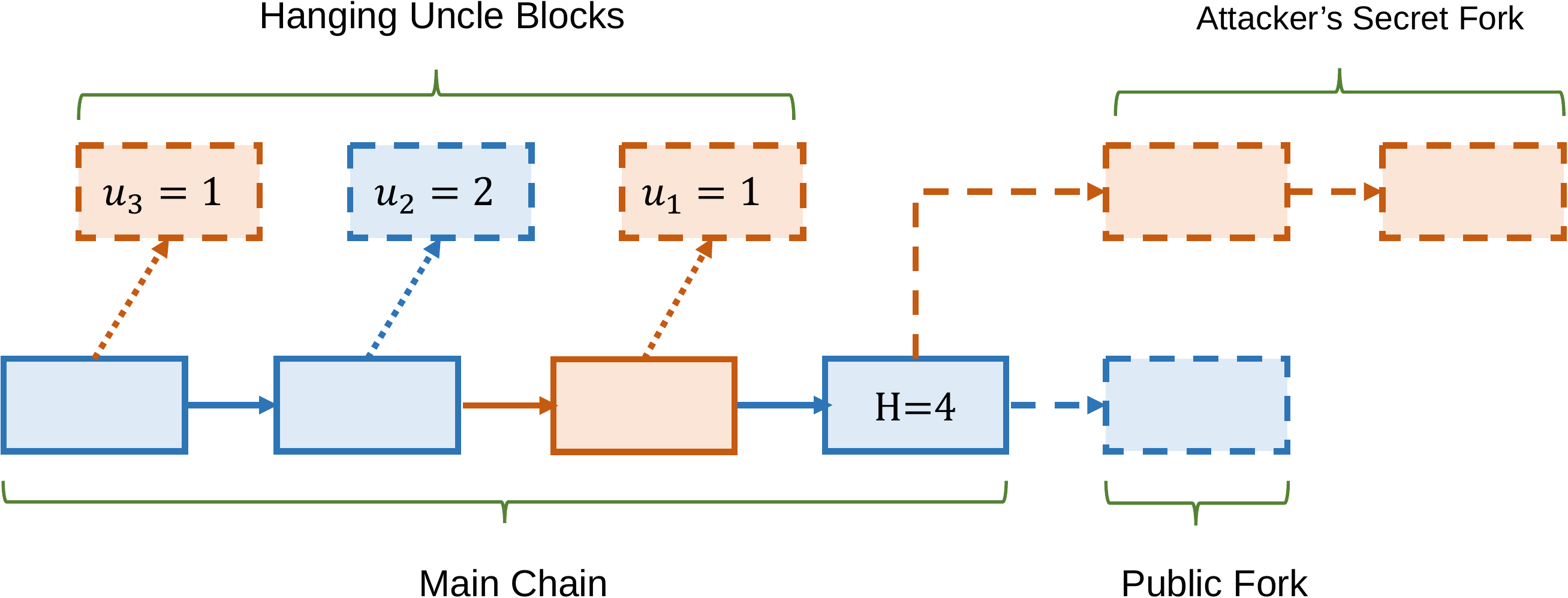}
		\caption{Ethereum state $(2, 1, {\tt irrelevant})$ and $U=\{1, 2, 1, 0, 0, 0\}$. Orange blocks are mined by the attacker and blue blocks  by the honest miner.}
		\label{fig:uncle1}
	\end{minipage}
	%
	\hfill
	\begin{minipage}[b]{\columnwidth}
		\centering
		\includegraphics[width=\textwidth]{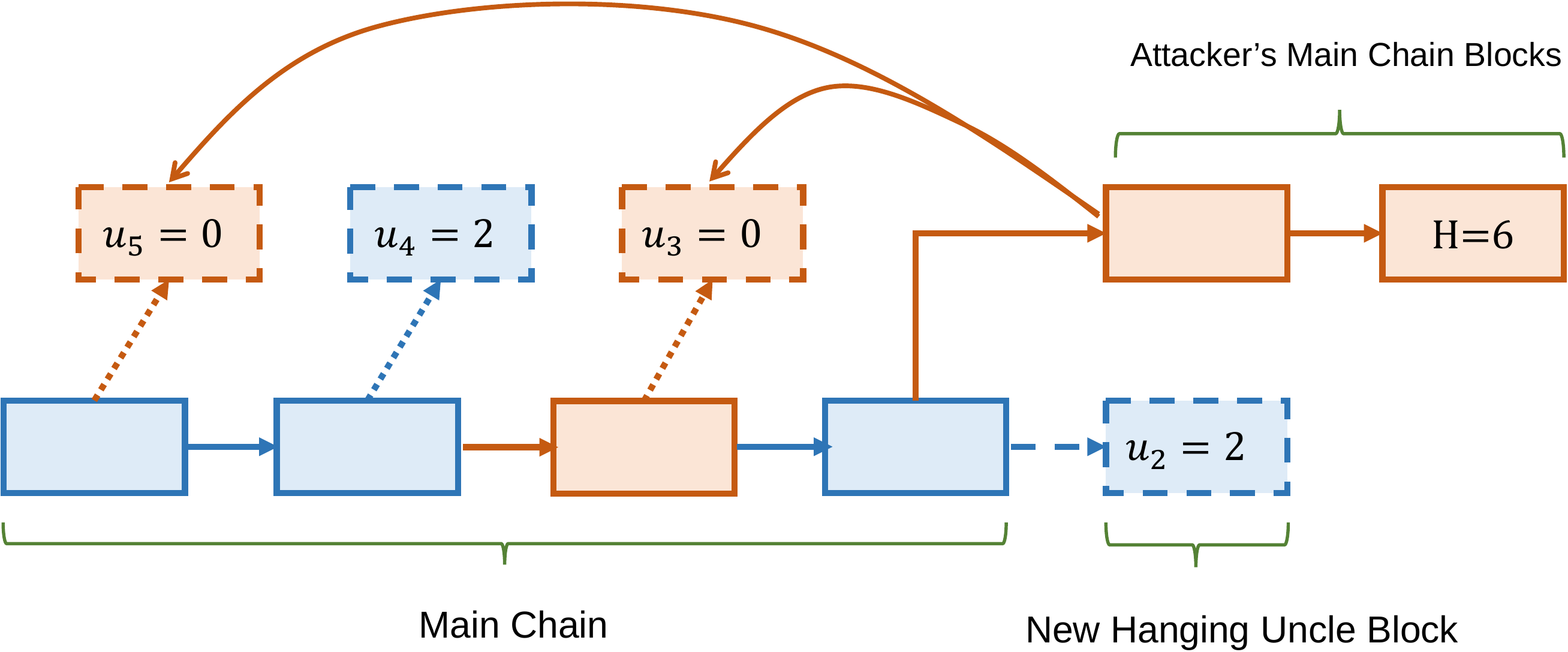}
		\caption{Ethereum state $(0, 0, {\tt irrelevant})$ and $U=\{0 ,2, 0, 2, 0, 0\}$. The attacker overrode the public fork with its secret fork and referred to two uncle blocks.}
		\label{fig:uncle2}
	\end{minipage}
\end{figure*}
\begin{figure}[t]
	\centering
	\includegraphics[width=0.8\columnwidth]{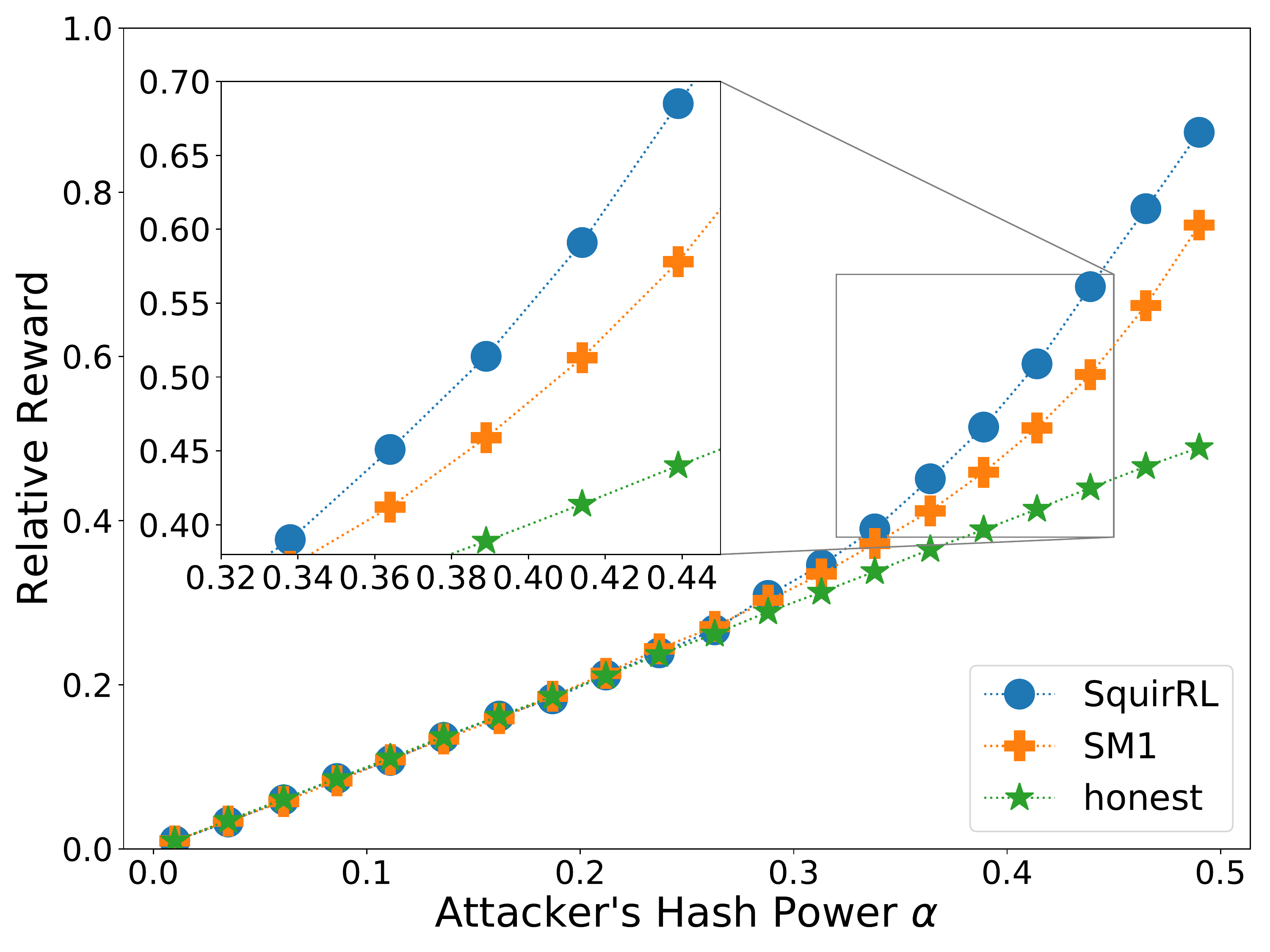}
	\caption{Ethereum relative reward as a function of  adversarial hash power. 
		RL beats the state-of-the-art schemes.
	}
	\label{fig:eth_r_vs_alpha}
\end{figure}
\subsection{Ethereum}
\label{sec:ethereum}
Our second experiment explores the Ethereum incentive mechanism.
In this setting, we were unable to recover the true optimal solution using an MDP solver, as the full Ethereum state space is too large.
Because of this, existing papers on selfish mining in Ethereum \protect{\cite{ritz2018impact,niu2019selfish,grunspan2019selfish}} do not derive an optimal solution like the one for Bitcoin \cite{sapirshtein2016optimal}.
This section illustrates how \name~can be used to explore the strategy space in scenarios where we do not have \emph{a priori} intuition about what strategies perform well and when MDP solvers are unable to recover meaningful results.

The Ethereum incentive mechanism is similar to Bitcoin's, except for its use of \emph{uncle rewards}. 
If a block is not a main-chain block but a child block of a main-chain block, it can be referenced as an {\em uncle block} (Figure \ref{fig:uncle1}). 
A block can have at most two uncle-block pointers and obtains $\frac{1}{32}$ of the full block reward for each. 
In addition, the miner of the uncle obtains a $\frac{8-k}{8}(1\le k \le 6)$-fraction of the full block reward, where $k$ is the height difference between the uncle block and the nephew block that points to it.

\smallskip
\noindent \textbf{Feature Extraction}. 
Here, we illustrate how to derive the feature extractor $\varphi$ for Ethereum.  Note that $L(C,T,E) = len(C)$ as noted in \Cref{sec:state-space} in the Ethereum example.  

To compute $\textrm{feat}(U(s))$, notice that a mined block can refer to (up to) \emph{any} two uncles in the 6-block history of the main chain. 
Hence, upon publishing $C$, the reward can depend on (a) the presence/absence of uncle blocks at each of the 6 prior main-chain blocks, and (b) who mined those uncle blocks. 
As such, we have $\sprof=(len(C_a), u)$, where $u \triangleq \{u_i\}_{i=1}^6$ encodes the information of the uncle blocks hanging on the main chain block of height $H-i$, where $H$ is the current height of the last common block of the main chain (Figure \ref{fig:uncle1}).
Each $u_i \in \{0, 1, 2\}$;
$u_i=0$ means there are no available uncle blocks at that height.
$u_i=1$ and $u_i=2$  mean that the uncle block was mined by the  attacker or honest miner, respectively. 
For instance, in  Figure \ref{fig:uncle1}, the attacker holds a secret fork with 2 blocks, while the public fork has only 1 block. 
The height of the main chain is $H=4$, and there are two uncle blocks mined by the attacker hanging from blocks of height $1$ and $3$. 
The uncle block mined by the honest miner is hanging at height $2$. So the uncle vector is $u=\{1, 2, 1, 0, 0, 0\}$.  

These uncle blocks as well as the $len(C)$ (as in Bitcoin) determine the instantaneous reward.  $len(C)$ is already included as part of the score, so we can leave out of the instananeous reward portion of our features.

Finally, Ethereum considers chains of equal length to be equally valid, regardless of when each were made public.  So there is no $\textrm{fork}$ feature required to include as part of $\textrm{act}(s)$, as all the other features will determine what actions are available.


Hence, our final features are $[len(C), H, u]$.
Notice that our framework for determining $\sprof$ does not directly store the uncle references; instead, it stores the minimum amount of information needed to \emph{compute} the reward for any given set of uncle references. 
This design choice prevents the state space from getting bloated.
For example, if we limit the maximum number of hidden blocks to 20, the state space size is around 291,600, which is out of range for many MDP solvers, but within range for DRL.
The uncles do not affect the state transitions.
Updates to vector $u$ caused by the addition of a new block have three effects: 
1) Any referred uncle blocks are removed from the vector by setting their corresponding entries to $0$; 
this prevents future blocks from referring to these already-referred uncle blocks. 
2) As the main chain's height is growing, the uncle indices are shifted, and any uncle blocks deeper than depth 6 are discarded, since they cannot be referred by any future blocks. 
3) Any fork shorter than the main chain is abandoned, and its first block becomes a new potential uncle block. 




\smallskip
\noindent \textbf{Performance.}
Figure \ref{fig:eth_r_vs_alpha} compares the relative rewards obtained by \name~to that of other selfish mining attacks in Ethereum~\protect{\cite{ritz2018impact,niu2019selfish,grunspan2019selfish}}. 
\name~outperforms these prior schemes, in part because those schemes implement  constrained strategies that are similar to SM1. 
In  fact, as Figure \ref{fig:eth_r_vs_alpha} also shows, Bitcoin OSM also outperforms prior works. Significantly here, \name~ at least matches the performance of OSM in most cases, and for hash power ranging from 25\% to 45\%,~outperforms OSM by 0.4\% to 1.0\%. We observe that~\name~implements a strategy that is more ``stubborn'' than OSM: it attempts to grow its secret fork more aggressively to compensate for the penalty of uncle rewards accruing to the honest player when the attacker fails to overwrite the main chain.

\subsection{Stochastic Hash Power in Real Cryptocurrencies}
\label{app:hash_power}
Here, we provide the visualization results of the total hash power and the relative hash power of the attacker for the Bitcoin, Monacoin, Dogecoin and Vertcoin tests in Sec.~\ref{sec:stoch_alpha}. 
\begin{figure*}[t!]
    \begin{minipage}[b]{0.32\linewidth}
    \centering
    \includegraphics[width=\columnwidth]{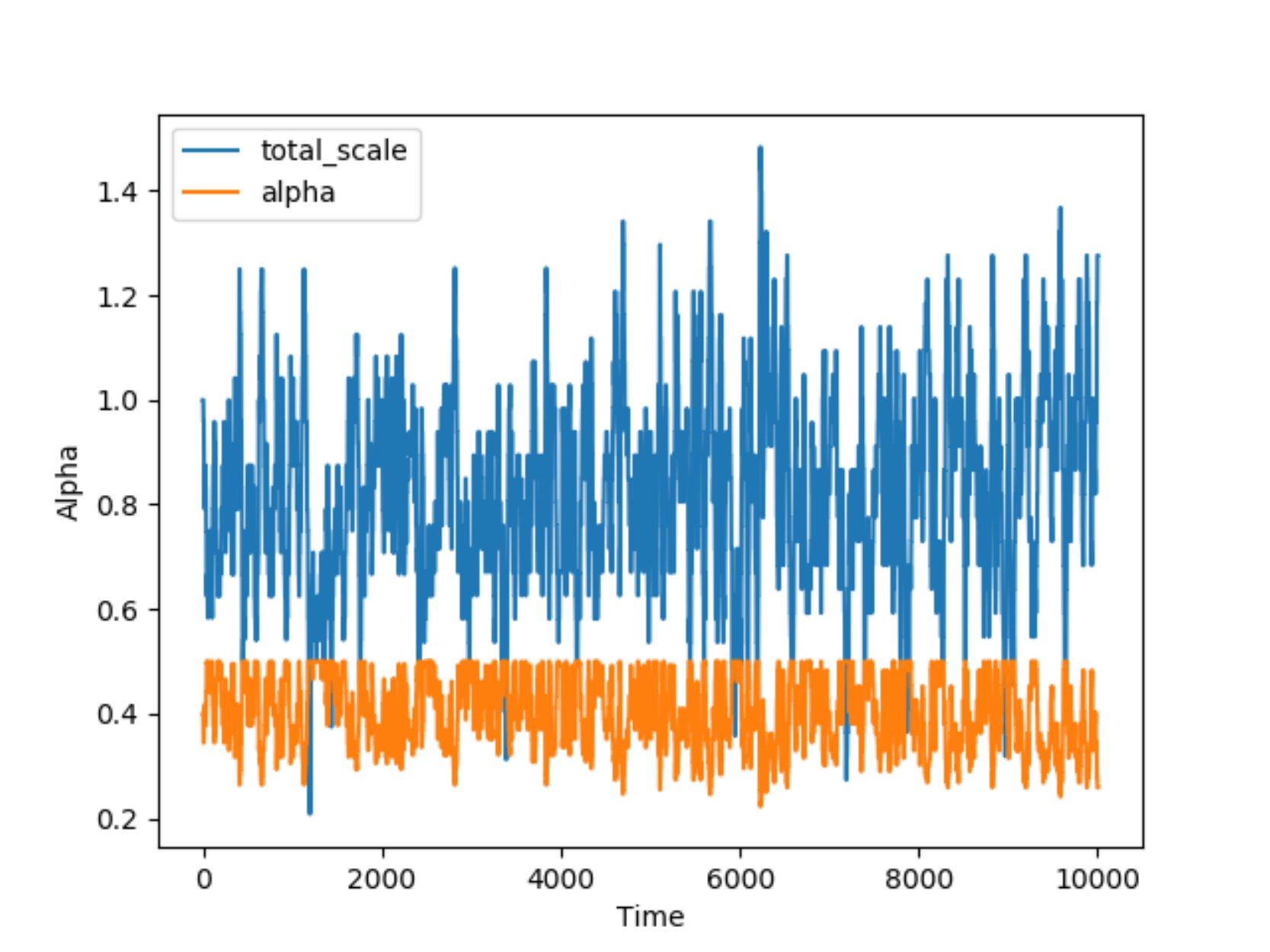}
    \caption{The total hash rate fluctuation(normalized) and the relative hash power for the attacker with initial $\alpha=0.4$ in Bitcoin from Sep. 2018 to Oct. 2018.}
        \label{fig:btc_real}
    \end{minipage}
    ~~
    \begin{minipage}[b]{0.32\linewidth}
    \centering
    \includegraphics[width=\columnwidth]{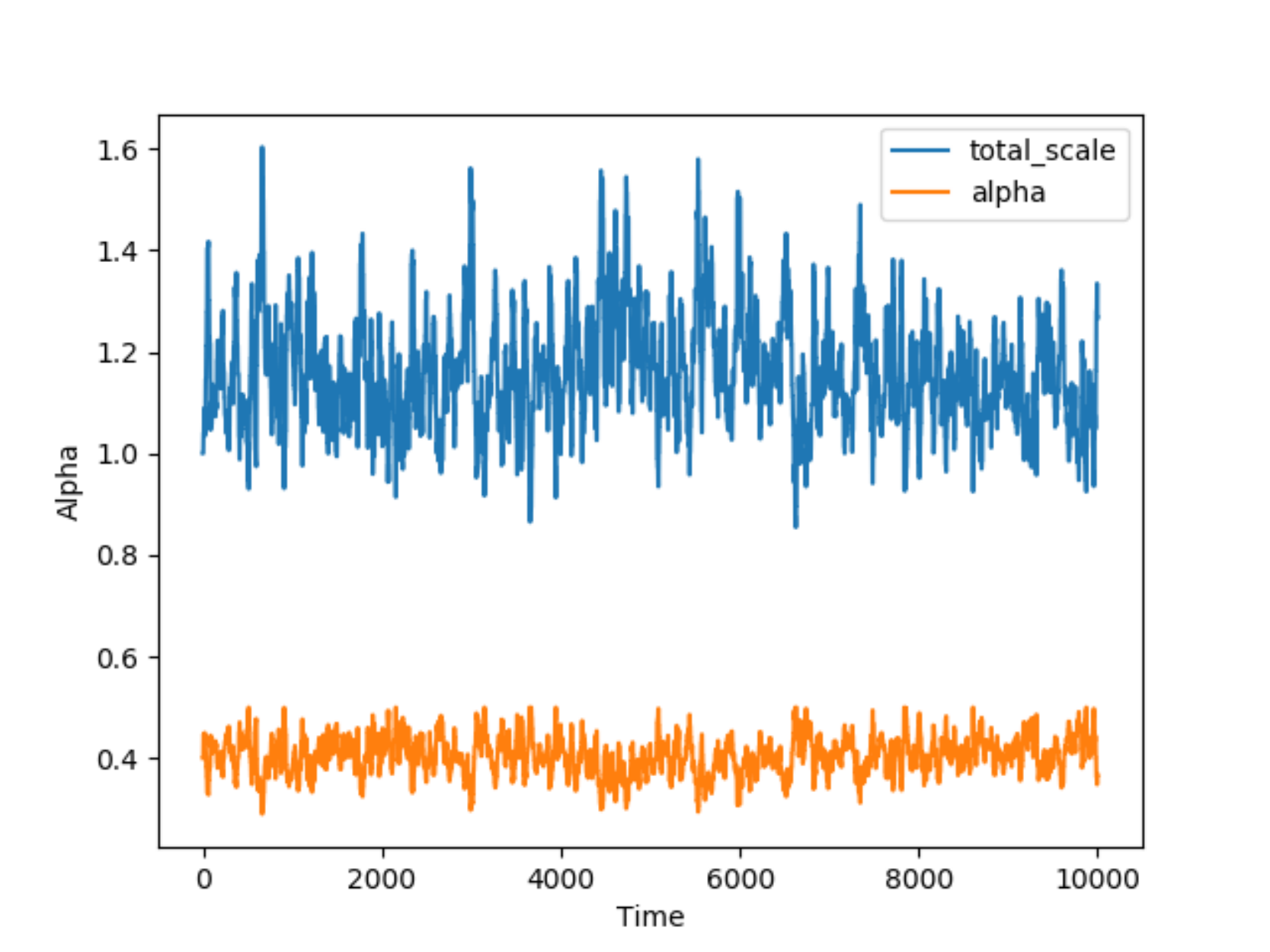}
    \caption{The total hash rate fluctuation(normalized) and the relative hash power for the attacker with initial $\alpha=0.4$ in Monacoin from Sep. 2018 to Oct. 2018.}
        \label{fig:Mona_real}
    \end{minipage}
    ~~
    \begin{minipage}[b]{0.32\linewidth}
    \centering
    \includegraphics[width=\columnwidth]{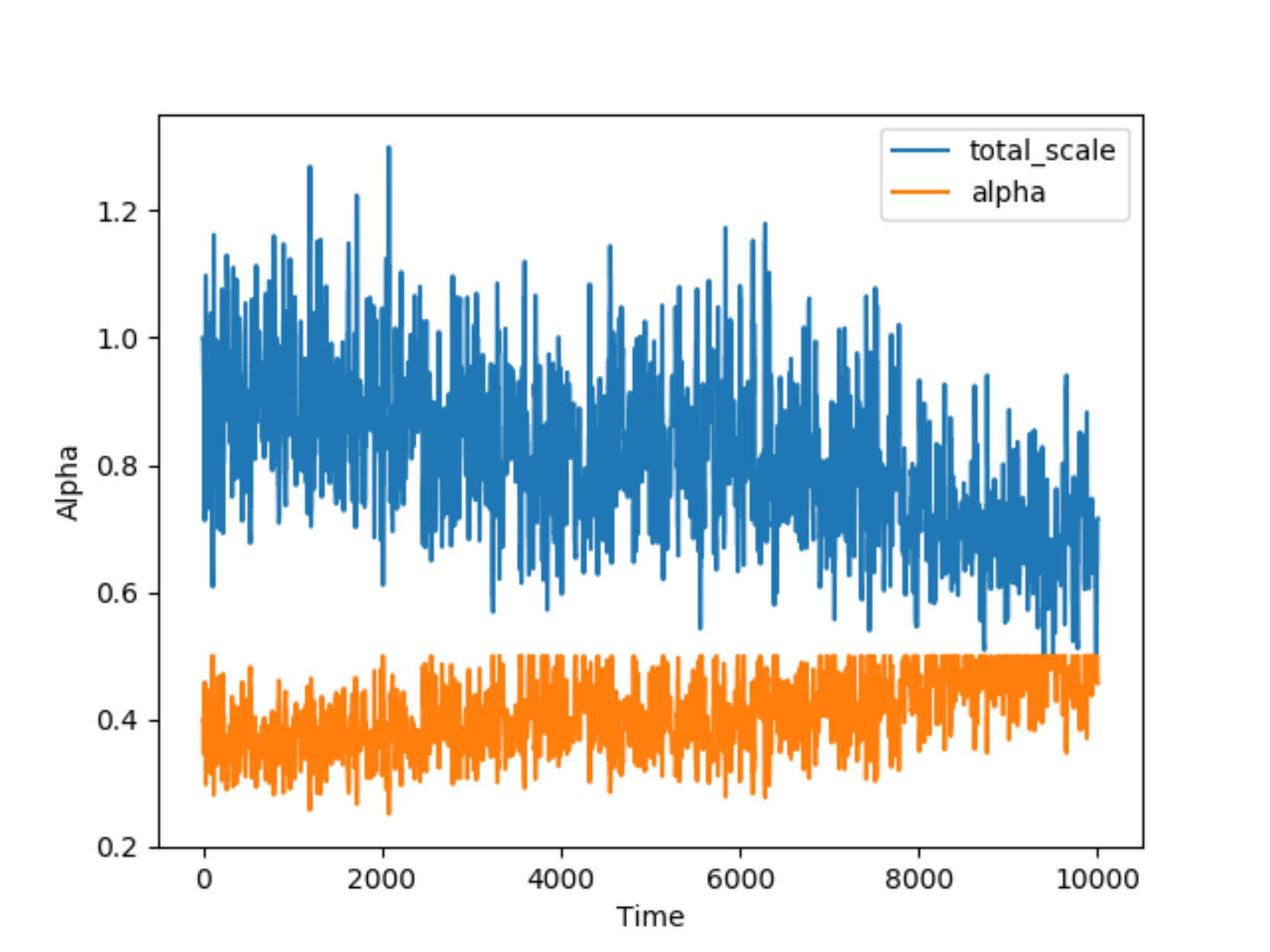}
    \caption{The total hash rate fluctuation(normalized) and the relative hash power for the attacker with initial $\alpha=0.4$ in Dogecoin from Sep. 2018 to Oct. 2018.}
        \label{fig:doge_real}
    \end{minipage}
    \end{figure*}
\begin{figure*}[t!]
\begin{minipage}[b]{0.32\linewidth}
\centering
\includegraphics[width=\columnwidth]{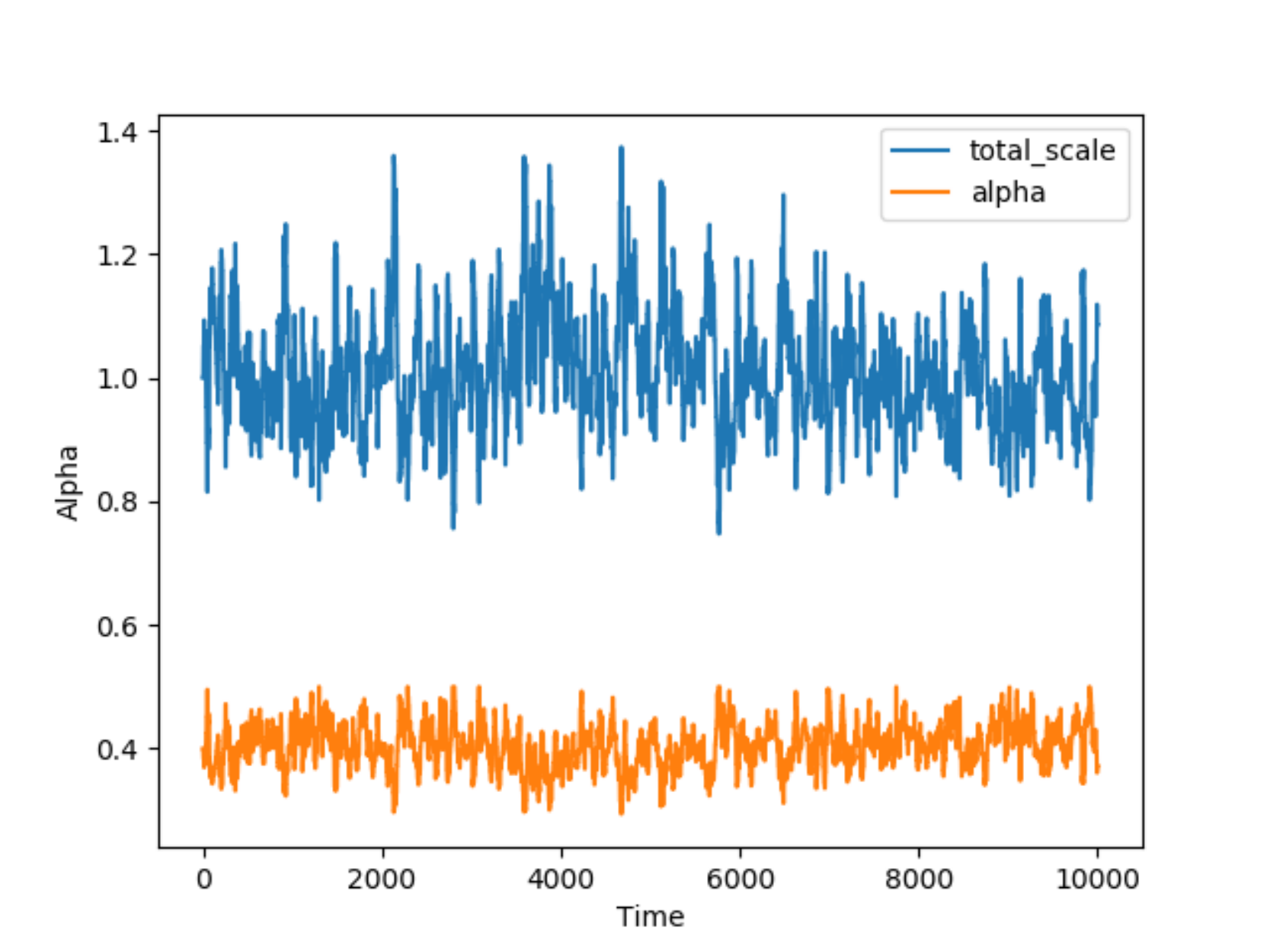}
\caption{The total hash rate fluctuation(normalized) and the relative hash power for the attacker with initial $\alpha=0.4$ in Vertcoin from Sep. 2018 to Oct. 2018.}
    \label{fig:vert_real}
\end{minipage}
\end{figure*}

\begin{table*}
	\centering
	\begin{tabular}{|l|l|l|}
		\hline
		& Multiplicative Reward & Additive Reward \\ \hline
		Justified,  Correct Vote  & $D^+_v=(1+m\rho/2)D_v$                   & $r_v=m\rho/2 D_v$          \\ \hline
		Justified, Wrong Vote     & $D^+_v=(1+m\rho/2)/(1+\rho)D_v$                 & $r_v=((1+m\rho/2)/(1+\rho)-1)D_v$          \\ \hline
		Unjustified, Correct Vote & $D^+_v=D_v$                 & $r_v=0$           \\ \hline
		Unjustified, Wrong Vote   & $D^+_v=1/(1+\rho)D_v$                & $r_v=-\rho/(1+\rho)D_v$           \\ \hline
	\end{tabular}
	\label{tab:casper_reward2}
	\caption{Reward rule in Casper FFG. The original multiplicative rule is not well-suited to RL systems, because the reward can become infinite over time, causing the value function to be ill-defined. Since the growth factor is small in practice, we choose an additive reward in our experiment that approximates the multiplicative reward over a finite time horizon. $D_v$ is the deposit of voter $v$ and $r_v$ is the immediate additive reward. $m$ is the fraction of correct votes.}
\end{table*}

\subsection{Non-monotonicity in OSM experiments}
\label{app:multi-agent-adaptive}
\label{app:adaptive_multiagent}
\begin{figure}[t]
	\centering
	\includegraphics[width=0.9\columnwidth]{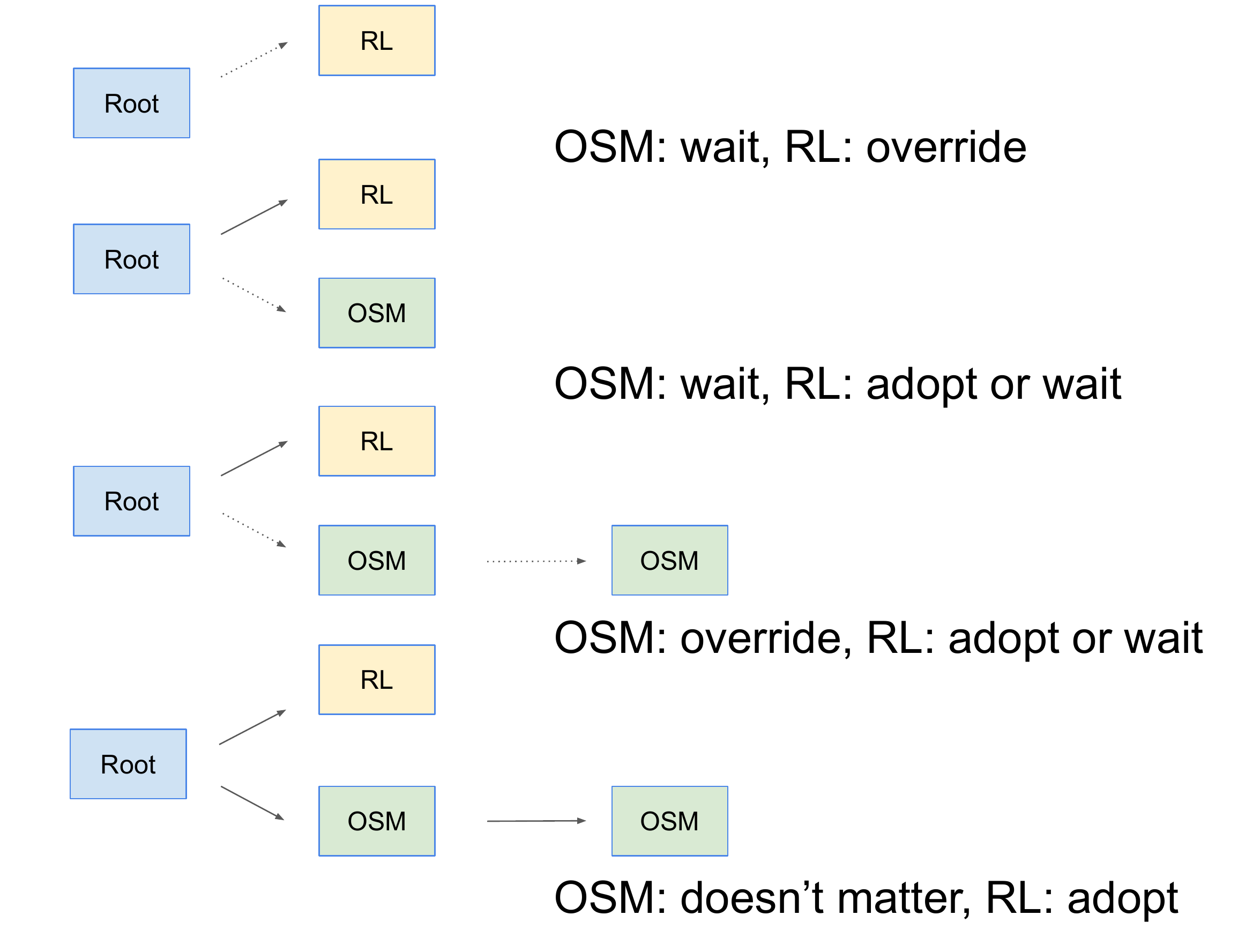}
	\caption{A sample trajectory. A block labeled "OSM" was mined by the OSM agent, and a block labeled "RL" was mined by the RL agent.  A dotted arrow means that portion of the chain is private.  The text after each block setting represents the actions that each respective agent chooses. Despite both agents practicing what would be honest mining in the single agent setting, blocks still get overwritten.}
	\label{fig:trace}
\end{figure}
\begin{figure}[t]
	\centering
	\includegraphics[width=0.8\columnwidth]{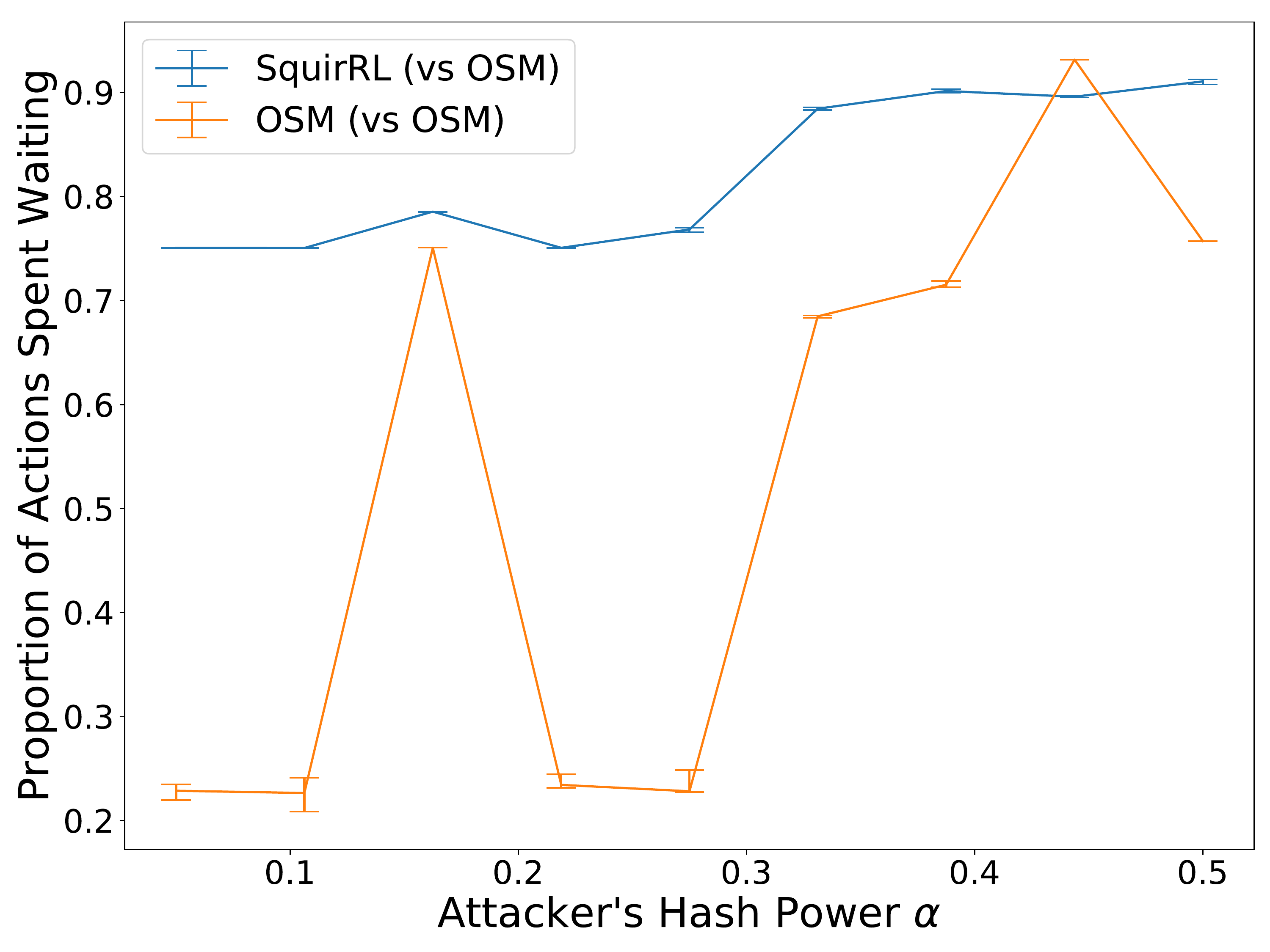}
	\caption{Agent C, when using~\name~, tends to use the "wait" action more than Agent B, who is following OSM.}
	\label{fig:osm_vs_rl_wait}
\end{figure}
In this section, we give more details on the cause of the non-monotonicity in Figures \ref{fig:osm_vs_rl_1} and \ref{fig:osm_vs_rl}.  As mentioned in Subsection \ref{sec:osm-ne}, there can be many OSM strategies that give the same reward in the single-agent setting, but give different rewards in the multi-agent setting.  We show a sample trajectory of the OSM strategy that caused the negative excess relative rewards of Figure \ref{fig:osm_vs_rl_1} and the positive non-monotonicity of Figure \ref{fig:osm_vs_rl_1} in Figure \ref{fig:trace}.

The critical difference between this strategy, which is honest in the single-strategic-agent setting and not honest in the multi-strategic-agent setting, with an overall honest strategy is the action the agent chooses at $(a,h,\textrm{fork}) = (0,0,.)$.  In the overall honest strategy, one should adopt at $(a,h,\textrm{fork}) = (0,0,.)$.  However, in the strategy depicted in Figure \ref{fig:trace}, the agent chooses to wait at $(a,h,\textrm{fork}) = (0,0,.)$.  In Figure \ref{fig:osm_vs_rl_wait}, we see that the non-monotonicity in strategy corresponds to the non-monotonicity in rewards we observed in Figures \ref{fig:osm_vs_rl_1} and \ref{fig:osm_vs_rl}.

Notably, this phenomenon wasn't observed in the multi-agent selfish mining analysis of \cite{semiselfish}.  This is because their semi-selfish agents automatically adopted at $(a,h,\textrm{fork}) = (0,0,.)$. 

\subsection{Details of the Casper FFG Experiment}
\label{sec:casper_experiment}
In our experiment, we use 10 as the epoch length that the voting events happen more frequently instead of 50. The voting probability $p_{vote}$ is set to $0.9$. The distribution $D_{vote}$ for the amount of votes cast per step is $\mathcal N(0.1, 0.05)$.

Casper FFG uses a multiplicative reward mechanism. 
If an agent acts according to protocol, her deposit is multiplied by a factor greater than 1, whereas if she disobeys protocol, her reward is multiplied by a factor less than 1. 
Table \ref{tab:casper_reward2} lists the precise formula for reward allocation under the column `Multiplicative Reward'.

We carefully pick the reward parameters to reflect the parameter combination deployed in practice. Firstly, the target total deposit pool is $D=10^7$. In the original setting, the parameter $\rho$ is calculated in every epoch by $\rho_i=\gamma D_i^{-p}+\beta (\text{ESF}_i-2)$, where $D_i$ is the deposit pool in epoch $i$ and $\gamma=7\cdot 10^{-3},\beta=2\cdot 10^{-7}, p=1/2$. Notice that the first term of the equation dominates the second term; therefore we omit the $\beta (\text{ESF}_i-2)$ term for simplicity. Therefore, we have a constant $\rho=2.21\cdot 10^{-6}$. Secondly, the original reward system is multiplicative, but we realize the total deposit pool is very big and the multiplicative factors are within $[1/(1+\rho), 1+1/2\cdot \rho]$, which is very close to 1. Within 1000 epochs, the absolute reward varies less than 1\%. Therefore, we simply fix the total deposit as $D=10^7$ and represent the rewards as additive rewards, described in table  \ref{tab:casper_reward2}, column `Additive Reward'.

We justify our parameters by numerical calculation. After the Thirdening upgrading, the production of ETH from block mining per year is around $4.9\cdot 10^6$ ETH. From \cite{buterin2019incentives}, the target deposit pool is $10^7$ and the in ideal situation, the annual interest of voting is 5\%, therefore the ETH production from voting reward is $5\cdot 10^5$ every year. Hence, the ratio between mining reward and voting reward is around $10:1$. In our experiment, one epoch contains 10 blocks, which means there are 20 ETH mined in an epoch. In an ideal situation voting round, the absolute reward for all voters is $\rho/2 D\approx 10.6$ ETH. We also need to divide the reward by 5, since we are using 10 as the epoch length instead of 50. Hence, the voting reward in one epoch is $2.12$ ETH, which matches the $10:1$ ratio.

\end{document}